\documentclass[twocolumn,trackchanges]{aastex62}

\usepackage{color}
\usepackage{natbib}
\usepackage{amsmath}
\usepackage{comment}

\newcommand{\nue}{\nu_{\rm e}}
\newcommand{\nueb}{\bar{\nu_{\rm e}}}
\newcommand{\nux}{\nu_x}
\newcommand{\ye}{Y_{\rm e}}

\newcommand{\pa}{\partial}

\newcommand{\rd}{{\rm d}}

\graphicspath{{./}{figures/}}

\received{\today}
\revised{\today}
\accepted{\today}
\submitjournal{ApJ}

\shorttitle{The Boltzmann Supernova Simulations with the Different EOSs}
\shortauthors{Harada et al.}

\begin{document}

\title{The Boltzmann-radiation-hydrodynamics Simulations of Core-collapse Supernovae with Different Equations of State: The Role of Nuclear Composition and the Behavior of Neutrinos}

\correspondingauthor{Akira Harada}
\email{harada@icrr.u-tokyo.ac.jp}

\author[0000-0003-1409-0695]{Akira Harada}
\affil{Institute for Cosmic Ray Research, University of Tokyo, 5-1-5 Kashiwanoha, Kashiwa, Chiba 277-8582, Japan}

\author{Hiroki Nagakura}
\affiliation{Department of Astrophysical Sciences, Princeton University, Princeton, NJ 08544, USA}

\author{Wakana Iwakami}
\affiliation{Advanced Research Institute for Science and Engineering, Waseda University, 3-4-1 Okubo, Shinjuku, Tokyo 169-8555, Japan}

\author{Hirotada Okawa}
\affiliation{Waseda Institute for Advanced Study, Waseda University, 1-6-1 Nishi-Waseda, Shinjuku-ku, Tokyo, 169-8050, Japan}

\author{Shun Furusawa}
\affiliation{Department of Physics, Tokyo University of Science, Shinjuku, Tokyo, 162-8601, Japan}
\affiliation{Interdisciplinary Theoretical and Mathematical Sciences Program (iTHEMS), RIKEN, Wako, Saitama 351-0198, Japan}

\author{Kohsuke Sumiyoshi}
\affiliation{Numazu College of Technology, Ooka 3600, Numazu, Shizuoka 410-8501, Japan}

\author{Hideo Matsufuru}
\affiliation{High Energy Accelerator Research Organization, 1-1 Oho, Tsukuba, Ibaraki 305-0801, Japan}

\author{Shoichi Yamada}
\affiliation{Advanced Research Institute for Science and Engineering, Waseda University, 3-4-1 Okubo, Shinjuku, Tokyo 169-8555, Japan}



\begin{abstract}
Using the Boltzmann-radiation-hydrodynamics code, which solves the Boltzmann equation for neutrino transport, we present the results of the simulations with the nuclear equations of state (EOSs) of Lattimer and Swesty (LS) and Furusawa and Shen (FS). We extend the simulation time of the LS model and conduct thorough investigations, though our previous paper briefly reported some of the results. Only the LS model shows the shock revival. This seems to originate from the nuclear composition: the different nuclear composition results in the different energy loss by photodissociation and hence the different strength of the prompt convection and the later neutrino-driven convection. The protoneutron star seen in the FS model is more compact than that in the LS model because the existence of multinuclear species softens the EOS. For the behavior of neutrinos, we examined the flux and the Eddington tensor of neutrinos. In the optically thick region, the diffusion of neutrinos and the dragging by the motion of matter determine the flux. In the optically thin region, the free-streaming determines it. The Eddington tensor is compared with that obtained from the M1-closure relation. The M1-closure scheme overestimates the contribution from the velocity-dependent terms in the semitransparent region.
\end{abstract}



\section{Introduction} \label{sec:intro}

Core-collapse supernovae (CCSNe) are considered to be the explosive death of massive stars. The energy source of the explosion is the gravitational energy released when the stellar iron core collapses to form a neutron star \citep[NS;][]{1934PNAS...20..254B}. It amounts to $\sim10^{53}\,{\rm erg}$. The CCSNe produce some heavy elements in the universe. In addition, the recent discovery and understanding of the binary NS merger imply that some $r$-process elements are produced there \citep{2017ApJ...848L..12A, 2017PhRvL.119p1101A, 2017PASJ...69..102T}. Therefore, it is important to understand CCSNe to explain the elemental evolution in the universe.

The standard scenario for the explosion is the neutrino heating mechanism \citep[see][for a review]{2012ARNPS..62..407J}. The stellar iron core eventually collapses. When the central density goes beyond the nuclear density, the matter suddenly becomes stiff, and the bounce shock is formed. This bounce shock propagates outward, consuming its energy by photodissociating the accreting heavy nuclei such as iron, and it finally stops propagation. SN modelers have been investigating how to revive this stalled shock. The leading hypothetical mechanism for shock revival is the neutrino heating mechanism \citep{1985nuas.conf..422W}. After the core bounce, a central hydrostatic object called a protoneutron star (PNS) is formed. The PNS is still hot and contains a lot of protons. The energy and the lepton number of the PNS are carried away by the neutrinos to evolve into the NS. In the neutrino heating mechanism, these emitted neutrinos are absorbed by matter just behind the shock and heat up the matter. Then, the shock restarts its propagation.

The progress of CCSN simulations started with 1D spherical symmetry. From the first stellar core-collapse simulation by \citet{1960PhRvL...5..235C}, CCSN simulations have been continuously improved. The works of \citet{2001PhRvD..63j3004L} and \citet{2005ApJ...629..922S} employed the very sophisticated simulation codes, which solve the Boltzmann equation for neutrino transport with general relativity \citep{1993ApJ...405..669M, 2001PhRvD..63j4003L, 1997ApJ...475..720Y, 1999A&A...344..533Y}. These sophisticated simulations did not show a successful shock revival. Nowadays, it is concluded that spherically symmetric CCSNe do not explode except for a special progenitor \citep{2006A&A...450..345K}.

From the middle of the 1990s, 2D axisymmetric simulations have been performed \citep{1994ApJ...435..339H, 1995ApJ...450..830B, 2003PhRvL..90x1101B, 2006A&A...447.1049B, 2009ApJ...694..664M, 2012ApJ...756...84M, 2013ApJ...767L...6B, 2015ApJ...800...10D, 2016ApJ...825....6S, 2016ApJ...818..123B, 2018ApJ...854...63O}. On this stage, shock revivals are found in simulations. These simulations revealed that the multidimensional effects such as turbulence help the neutrino heating and shock revival. 

The Boltzmann neutrino transport, which is employed in, e.g., \citet{1999A&A...344..533Y, 2000ApJ...539..865B, 2001PhRvD..63j4003L} under spherical symmetry, requires a lot of numerical resources; thus, almost all multidimensional simulations so far use approximate neutrino transport. For example, the flux-limited diffusion scheme \citep{2006ApJ...640..878B, 2007PhR...442...23B}, two-moment scheme (M1 closure: \citet{2012ApJ...755...11K, 2015MNRAS.453.3386J}, variable Eddington factor: \citet{2002A&A...396..361R}), and isotropic diffusion source approximation \citep{2009ApJ...698.1174L} are utilized. In addition, the ray-by-ray(-plus) approach \citep{2002A&A...396..361R, 2006A&A...447.1049B} is sometimes used. As the employed approximation schemes are different among SN modelers, the outcomes such as the explodability and explosion energy of the simulations are also different. Some collaboration works have reported on a code comparison project, and they show basic agreement in 1D simulations or in the early stages of the postbounce dynamics in multiple dimensions \citep{1999A&A...344..533Y, 2005ApJ...620..840L, 2010ApJS..189..104M, 2016ApJ...831...81S, 2018A&A...619A.118C, 2018MNRAS.481.4786J, 2019ApJ...873...45G, 2018JPhG...45j4001O, 2019JPhG...46a4001P}. They guarantee that the basic ingredients are correctly included in the codes, but the origin of some differences is yet to be answered.

Even though the shock revives in 2D simulations, there are still some problems. The simulated explosion energy is about $10^{50}\,{\rm erg}$, at least at several hundreds of milliseconds after the core bounce. This is much smaller than the observed value of $10^{51}\,{\rm erg}$. Current development of computational resources allows SN modelers to perform long-term simulations exceeding $1\,{\rm s}$ after the core bounce, and some of these simulations indicate that the continued evolution of the explosion energy may reach the observed explosion energy \citep{2013ApJ...767L...6B, 2016ApJ...818..123B}. On the other hand, such long-term and relatively slow heating may be problematic in explaining another observable, the amount of ${}^{56}{\rm Ni}$. \citet{2019MNRAS.483.3607S} suggested from the analytic model calibrated by numerical simulations that the observed amount of ${}^{56}{\rm Ni}$ is reproduced if the heating rate exceeds $\mathcal{O}(10^{51})\,{\rm erg\,s^{-1}}$, about $10$ times larger than what is expected in simulations. Therefore, the problem related to the explosion energy, or the heating rate, still remains.

Recent SN modelers are going to 3D simulations, thanks to increasing computational power \citep{2002ApJ...574L..65F, 2012ApJ...749...98T, 2013ApJ...770...66H, 2014ApJ...786...83T, 2013PhRvL.111l1104T, 2014PhRvD..90d5032T, 2015ApJ...807L..31L, 2015ApJ...801L..24M, 2015MNRAS.453..287M, 2016ApJS..222...20K, 2018ApJ...855L...3O, 2019MNRAS.482..351V}. This is because nature is, of course, 3D, and hence 3D simulations are appropriate for comparing to observations such as the anisotropic distribution of ejected elements, the aspherical morphology of SN remnants, and so on \citep{2017ApJ...842...13W, 2020ApJ...888..111O}. However, although some simulations show shock revival, problems similar to those in 2D simulations still remain.

Using SN simulations, roles of microphysics such as nuclear equations of state (EOSs) are investigated. Various nuclear EOS models have been developed for SN simulations \citep[e.g.][]{1991NuPhA.535..331L, 1998NuPhA.637..435S, 2011ApJ...738..178F, 2013ApJ...772...95F, 2017NuPhA.957..188F, 2017JPhG...44i4001F, 2017NuPhA.961...78T, 2019ApJ...887..110S}. The effects of these EOSs on the SN dynamics have been discussed \citep[e.g.,][]{2009A&A...496..475M, 2013ApJ...765...29C, 2013ApJ...764...99S, 2014EPJA...50...46F, 2019PhRvC.100e5802S}; those on the multimessenger signals, i.e., neutrinos and gravitational waves, have been examined \citep[e.g.,][]{2018ApJ...857...13P, 2020ApJ...891..156N}. These results imply that the soft EOSs such as the Lattimer--Swesty EOS \citep{1991NuPhA.535..331L} tend to show robust shock revival.

We should note that the failure to revive shocks in 1D simulations is concluded from the general relativistic Boltzmann-radiation-hydrodynamics simulations. By omitting the approximations in the governing equations, we can robustly assess the effects of the uncertainty in the input physics: the nuclear EOSs and microphysical reactions. Because there still remain some problems in 2D simulations, it is important to use first-principles simulations, which do not use approximations for the governing equations, to understand the origin of the problem.

We utilized the improved computational resources to solve the Boltzmann-radiation hydrodynamics under axisymmetry, instead of performing 3D simulations. The development of the Boltzmann-radiation-hydrodynamics code is reported in \citet{2012ApJS..199...17S, 2014ApJS..214...16N, 2017ApJS..229...42N} and briefly explained in section \ref{sec:method}. We employed the $S_{\rm N}$ method for neutrino transport, and the comparison with Monte Carlo transport is presented in \citet{2017ApJ...847..133R}. Furthermore, \cite{2020MNRAS.496.2000C} also developed a multidimensional Boltzmann-neutrino-transport solver, indicating the necessity of the exact treatment of neutrino transport.

We report the results of thorough analyses of the simulations using the Boltzmann-radiation-hydrodynamics code with different nuclear EOSs in this paper. We have reported the results of the simulations with the code in \citet{2018ApJ...854..136N, 2019ApJ...872..181H} so far. The initial work, \citet{2018ApJ...854..136N}, demonstrated the novel features of our Boltzmann-radiation-hydrodynamics code with simulations with different nuclear EOSs: the differences in the dynamical features were discussed briefly; the momentum space distributions of neutrinos were deeply analyzed, but only limited time snapshots and spatial points were considered. Therefore, thorough analyses of the dynamical features and neutrino distributions at various times and spatial regions are necessary; we analyzed the effects of the different EOSs and the behavior of the neutrinos deeply and widely.

The simulation time is extended to confirm the fate of the shock. In the previous paper, the simulation time was not enough to see whether the shock is revived or not. Thus, we ran additional simulation continuing from the previous paper and show the results here. Thanks to the extended simulation, the shock revival is more noticeable than in the previous paper.

This paper is structured as follows. In section \ref{sec:method}, a summary of our code and the models are presented. Next, in section \ref{sec:dynamics}, we discuss the comparison of the postbounce hydrodynamics and some observable neutrino quantities from the viewpoint of the EOSs. Then, in section \ref{sec:neutrinos}, we give detailed analyses of the neutrino quantities: the neutrino fluxes and the Eddington tensors. Finally, we conclude our paper in section \ref{sec:concl}. The units with $c=G=\hbar=1$ are considered in this paper unless otherwise stated. The signature of the spacetime metric is $(-+++)$. Greek and Latin indices run over $0$--$3$ for spacetime and $1$--$3$ for space, respectively.

\section{Numerical Modeling} \label{sec:method}
The Boltzmann-radiation-hydrodynamics code used in this paper solves the directly discretized Boltzmann equation (the so-called $S_N$ method). The details of the code are described in \citet{2012ApJS..199...17S, 2014ApJS..214...16N, 2017ApJS..229...42N}.
The Boltzmann equation for neutrinos with a general metric is cast in the conservative form \citep{PhysRevD.89.084073}:
\begin{eqnarray}
\frac{1}{\sqrt{-g}}\frac{\partial}{\partial x^\alpha}\Bigg|_{q_i} \left[ \left( e_{(0)}^\alpha + \sum_{i=1}^3 \ell_{(i)} e_i^\alpha \right) \sqrt{-g}f\right] && \nonumber\\
- \frac{1}{\epsilon^2}\frac{\partial}{\partial \epsilon}(\epsilon^3 f \omega_{(0)}) + \frac{1}{\sin \theta_\nu} \frac{\partial}{\partial \theta_\nu}(\sin \theta_\nu f \omega_{(\theta_\nu)} ) && \nonumber\\
+ \frac{1}{\sin^2 \theta_\nu}\frac{\partial}{\partial \phi_\nu}(f \omega_{(\phi_\nu)}) = S_{\rm rad},
\end{eqnarray}
\begin{eqnarray}
\ell_{(i)} &:=& (\cos \theta_\nu, \sin \theta_\nu \cos \phi_\nu, \sin \theta_\nu \sin \phi_\nu), \\
\omega_{(0)} &:=& \epsilon^{-2}p^\alpha p_\beta \nabla_\alpha e_{(0)}^\beta, \\
\omega_{(\theta_\nu)} &:=& \sum_{i=1}^3 \omega_i \frac{\partial \ell_{(i)}}{\partial \theta_\nu}, \\
\omega_{(\phi_\nu)} &:=& \sum_{i=2}^3 \omega_i \frac{\partial \ell_{(i)}}{\partial \phi_\nu}, \\
\omega_i &:=& \epsilon^{-2} p^\alpha p_\beta \nabla_\alpha e_{(i)}^\beta,
\end{eqnarray}
where $x^\alpha$, $g$, $e^\alpha_{(\mu)}$, $f$, $p^\alpha$, $\epsilon:=-p_\mu e^\mu_{(0)}$, $\theta_\nu$, $\phi_\nu$, and $S_{\rm rad}$ are the spatial coordinates, the determinant of the metric $g_{\mu\nu}$, the tetrad bases, the distribution function, the four-momentum of the neutrino, the energy of the neutrino, the zenith and azimuthal angle of the neutrino flight direction with respect to the tetrad basis, and the collision term, respectively. The zeroth tetrad basis $e^\mu_{(0)}$ is chosen to be the normal vector to the spatial hypersurface $n^\mu = (\alpha^{-1}, -\alpha^{-1}\beta^i)$, where $\alpha$ and $\beta^i$ are the lapse function and the shift vector. According to the $3+1$ decomposition, the spacetime metric is decomposed as
\begin{eqnarray}
\rd s^2 &=& g_{\mu\nu} \rd x^\mu \rd x^\nu \\
&=& -\alpha^2 \rd t^2 + \gamma_{ij}(\rd x^i + \beta^i \rd t)(\rd x^j + \beta^j \rd t),
\end{eqnarray}
where $\gamma_{\mu\nu} = g_{\mu\nu} + n_\mu n_\nu$ is the spatial metric. Because we choose polar coordinates $(r,\theta,\phi)$, the other tetrad bases are chosen to be
\begin{eqnarray}
e_{(1)}^\alpha &=& \gamma_{rr}^{-1/2} \pa_r, \label{eq:tet1} \\
e_{(2)}^\alpha &=& -\frac{\gamma_{r\theta}}{\sqrt{\gamma_{rr}(\gamma_{rr}\gamma_{\theta\theta}-\gamma_{r\theta}^2)}}\pa_r + \sqrt{\frac{\gamma_{rr}}{\gamma_{rr}\gamma_{\theta\theta}-\gamma_{r\theta}^2}}\pa_\theta, \label{eq:tet2} \\
e_{(3)}^\alpha &=& \frac{\gamma^{r\phi}}{\sqrt{\gamma^{\phi\phi}}}\pa_r+\frac{\gamma^{\theta\phi}}{\sqrt{\gamma^{\phi\phi}}}\pa_\theta+\sqrt{\gamma^{\phi\phi}}\pa_\phi, \label{eq:tet3}
\end{eqnarray}
where $\pa_i$ are the coordinate bases of the vector. Although these expressions can be applied to the general, curved spacetime, we only focus on the flat spacetime with drifting coordinates. The coordinates move with the PNS for their centers to match. This is realized by the flat spacetime with the nonzero shift vector. This frame is nothing but an acceleration frame. The method to determine the shift vector is discussed in \citet{2017ApJS..229...42N}. Because the coordinate drift is not significant in the current models as discussed in section \ref{sec:overviewdyna}, we just call it the laboratory frame. Although this equation describes the time evolution of the six-dimensional distribution function, we impose axisymmetry due to limited computational resources, and as a consequence, the distribution function is a function in the five-dimensional phase space (two in configuration space and three in momentum space).

The neutrino reactions are based on the standard set \citep{1985ApJS...58..771B}: the (anti)electron capture on the nucleon and nuclei, the scattering off the nucleon and nuclei, and the pair neutrino production from the electron--positron pair. However, there are several modifications: the electron capture on heavy nuclei is updated according to \citet{2010NuPhA.848..454J, 2000NuPhA.673..481L, 2003PhRvL..90x1102L}; the inelastic scattering off electrons is incorporated; and nucleon--nucleon bremsstrahlung is implemented. Contrary to the recent work \citep{2019ApJ...880L..28N}, the electron capture on light nuclei is not incorporated. Instead, it is artificially replaced by the weak interactions with the free nucleons, which are the constituents of light nuclei. This treatment is similar to the DV112 model in \citet{2019ApJS..240...38N}.  We set the same rates for the reactions involving the heavy-lepton-type neutrinos $\nu_\mu$, $\bar{\nu_\mu}$, $\nu_\tau$, and $\bar{\nu_\tau}$ \citep[but see][]{2017PhRvL.119x2702B}. They are collectively denoted by $\nu_x$. Therefore, we solve the Boltzmann equations for $\nu_{\rm e}$, $\bar{\nu_{\rm e}}$, and $\nu_x$.

As for the hydrodynamics, we solve the Newtonian hydrodynamics equations in the acceleration frame:
\begin{equation}
\pa_t (\sqrt{-g}\rho) + \pa_i(\sqrt{-g}\rho v^i) = 0,
\end{equation}
\begin{eqnarray}
&&\pa_t (\sqrt{-g}\rho v_r) + \pa_i\left(\sqrt{-g}(\rho v_r v^i + p\delta^i_r)\right) = \nonumber \\
&& \sqrt{-g}\rho \left(-\pa_r \psi + r (v^\theta)^2 + r\sin^2 \theta (v^\phi)^2 + \frac{2p}{r\rho} \right) \nonumber \\
&&-\sqrt{-g}G_r - \sqrt{-g} \rho \dot{\beta_r},
\end{eqnarray}
\begin{eqnarray}
&&\pa_t (\sqrt{-g}\rho v_\theta) + \pa_i\left(\sqrt{-g}(\rho v_\theta v^i + p\delta^i_\theta)\right) = \nonumber \\
&& \sqrt{-g}\rho \left(-r^2 \pa_\theta \psi + \sin\theta \cos\theta (v^\phi)^2 + \frac{p\cos\theta}{\rho\sin\theta}\right) \nonumber \\
&& -\sqrt{-g}G_\theta - \sqrt{-g}\rho \dot{\beta_\theta}, \label{eq:hydrotheta}
\end{eqnarray}
\begin{eqnarray}
&&\pa_t (\sqrt{-g}\rho v_\phi) + \pa_i\left(\sqrt{-g}(\rho v_\phi v^i + p\delta^i_\phi)\right) = \nonumber \\
&&-\sqrt{-g}\rho\pa_\phi \psi - \sqrt{-g}G_\phi - \sqrt{-g}\rho \dot{\beta_\phi}, \label{eq:hydrophi}
\end{eqnarray}
\begin{eqnarray}
&&\pa_t \left(\sqrt{-g} (e + \frac{1}{2}\rho v^2)\right) + \pa_i\left(\sqrt{-g}(e + p + \frac{1}{2}\rho v^2)v^i \right) = \nonumber \\
&& -\sqrt{-g}\rho v^i \pa_i \psi -\sqrt{-g}G^t - \sqrt{-g}\rho v^i \dot{\beta_i},
\end{eqnarray}
and
\begin{equation}
\pa_t(\sqrt{-g}\rho Y_{\rm e} ) + \pa_i (\sqrt{-g}\rho Y_{\rm e}v^i) = -\sqrt{-g}(\Gamma_{\nu_{\rm e}} - \Gamma_{\bar{\nu_{\rm e}}}).
\end{equation}
The symbols $\rho$, $v^i$, $p$, $e$, $Y_{\rm e}$, and $\psi$ represent the density, the velocity, the pressure, the internal energy, the electron fraction, and the Newtonian gravitational potential, which obeys
\begin{equation}
\Delta \psi = 4\pi \rho,
\end{equation}
respectively. In these Newtonian hydrodynamics equations, we consider the flat spacetime metric, i.e., $\sqrt{-g} = r^2\sin\theta$. The exchange of energy and momentum between matter and neutrinos is
\begin{equation}
G^\mu = \int p^\mu \epsilon S_{\rm rad} \rd V_p, \label{eq:interaction}
\end{equation}
where $\rd V_p$ is the invariant volume element in the momentum space. The reaction rate of the electron-type neutrinos and antineutrinos are
\begin{equation}
\Gamma_{\nu_{\rm e}/\bar{\nu_{\rm e}}} = m_{\rm u} \int \epsilon S_{{\rm rad},\nu_{\rm e}/\bar{\nu_{\rm e}}} \rd V_p,
\end{equation}
where $m_{\rm u}$ and $S_{{\rm rad},\nu_{\rm e}/\bar{\nu_{\rm e}}}$ are the atomic mass unit and the collision term for the indicated neutrino species.

The EOSs adopted in this paper are the Furusawa--Shen \citep[FS;][]{2011ApJ...738..178F, 2013ApJ...772...95F} EOS and the Lattimer--Swesty \citep[LS;][]{1991NuPhA.535..331L} EOS. The FS EOS is based on the Shen EOS \citep{1998NuPhA.637..435S}, and nuclear statistical equilibrium (NSE) is considered for the ensemble of nuclei in order to calculate the thermodynamical and statistical properties of nonuniform matter. The Shen EOS, the basis of the FS EOS, models the strong interaction by the relativistic mean field theory with the TM1 parameter set. On the other hand, the LS EOS is based on the liquid drop model of the nuclei and the Skyrm-type interaction. As for the composition, \citet{1991NuPhA.535..331L} assume that the heavy nuclei are represented by a single nuclear species (single nuclear approximation, SNA), and only the alpha particle is considered as the light nuclei. Among the EOSs offered by \citet{1991NuPhA.535..331L}, we choose the EOS with the incompressibility parameter $K=220\,{\rm MeV}$. Because our simulation code uses tabulated EOSs, we convert the subroutine EOS originally provided by \citet{1991NuPhA.535..331L} to the tabulated EOS.\footnote{In the conversion process, the mass difference between the neutron and the proton is incorrectly treated. As a result, the neutrino cooling is slightly underestimated. This underestimation enhances the shock radius mildly, by a few percent at the prompt shock phase and less than $10\%$ at the shock stagnation phase according to 1D test simulations. This error is so small that the results presented in this paper do not change significantly.} For a given set of density, temperature, and electron fraction, the electron capture rates on the heavy nuclei are calculated based on the heavy nucleus abundance of the FS EOS. The same rate as a function of the thermodynamic quantities is used for both models with the FS and LS EOSs. Although the more sophisticated Furusawa--Togashi (FT) EOS is available \citep{2017NuPhA.961...78T, 2019ApJS..240...38N}, we do not employ it in this paper. Because the weak interaction rates are different between the EOSs in this paper and the FT EOS, we could not focus on the properties of nuclear matter if we compared the simulations with the LS/FS and FT EOSs.

The formulation described above is numerically evolved with the following schemes. The numerical flux of the Boltzmann equation is evaluated by the combinations of the upwind and central difference scheme according to the local mean free path (MFP). Specifically, we carefully discretize the Boltzmann equation to guarantee the steady-state infinite homogenous solution, i.e., a constant distribution function with respect to the spacetime and momentum. The equation is evolved semi-implicitly, thus we need to solve large linear coupled equations. The Bi-CGSTAB method \citep{Saad2003} is utilized for the matrix inversion with the weighted point Jacobi preconditioner. For the hydrodynamics equations, the Harten--Lax--van Leer (HLL) scheme \citep{1983SIAMR...25...35H} with piecewise-parabolic interpolation \citep{Colella1984174} determines the numerical flux. The second-order Runge--Kutta method is adopted for the time integration. The Poisson equation for the gravitational potential $\psi$ is solved by direct multiplication of the inverse matrix of the discretized Laplacian operator. The inverse matrix is calculated using the MICCG method \citep{2011ApJ...731...80N}.

We run the two-dimensional axisymmetric simulations from the beginning of the prompt convection. The progenitor model is the $11.2\,M_\odot$ model in \citet{2002RvMP...74.1015W}. Until the negative entropy gradient is formed, one-dimensional simulations are performed from the onset of the collapse. When the negative entropy gradient is formed at $t_{\rm pb}\sim 0.6\,{\rm ms}$, where $t_{\rm pb}$ is the postbounce time, the hydrodynamical quantities and the neutrino distributions are mapped to the two-dimensional simulations with seed perturbations of $0.1\%$ in the radial velocity inside the region $30 \le r \le 50\,{\rm km}$. Because the shock radius is $\sim 30\,{\rm km}$ at that time, this is the preshock velocity perturbation.  The computational domain is $0 \le r \le 5000\,{\rm km}$ and $0 \le \theta \le \pi$. The radial and zenith coordinates are divided into $384$ and $128$ grids, respectively. Although our code can solve equation (\ref{eq:hydrophi}) simultaneously \citep{2019ApJ...872..181H, 2020arXiv200402091I}, we do not solve it for the nonrotating and axially symmetric simulation in this paper. The neutrino energies range from $0$ to $300\,{\rm MeV}$ and are divided into $20$ grids. The $\theta_\nu$ and $\phi_\nu$ cover a full solid angle and are divided into $10$ and $6$ bins, respectively.

The LS and FS models are followed up to $400\,{\rm ms}$ and $300\,{\rm ms}$ after the core bounce, respectively. The LS model was followed until $300\,{\rm ms}$ in \citet{2018ApJ...854..136N}. The time was not enough to judge whether the shock is revived or not, and hence we extend the time for the LS model. We do not extend the time for the FS model because the FS model clearly fails to explode.

\section{Dynamical features} \label{sec:dynamics}
In this section, we discuss the dynamical features of our two simulations. First, we give an overview of the time evolutions of the shock waves, neutrino luminosities and energies, entropies, and the PNS motion in section \ref{sec:overviewdyna}. Next, we give a detailed analysis of the deformation of the shock and the convection in section \ref{sec:defshock}. In section \ref{sec:timescale}, we investigate the origin of the difference in the shock evolution between the LS and FS models using the timescale ratio. Then, we discuss the influence of the EOSs on the structure of the PNS in section \ref{sec:PNSstructure}.

\subsection{Overview of the dynamics} \label{sec:overviewdyna}

\begin{figure*}[ht!]
\plotone{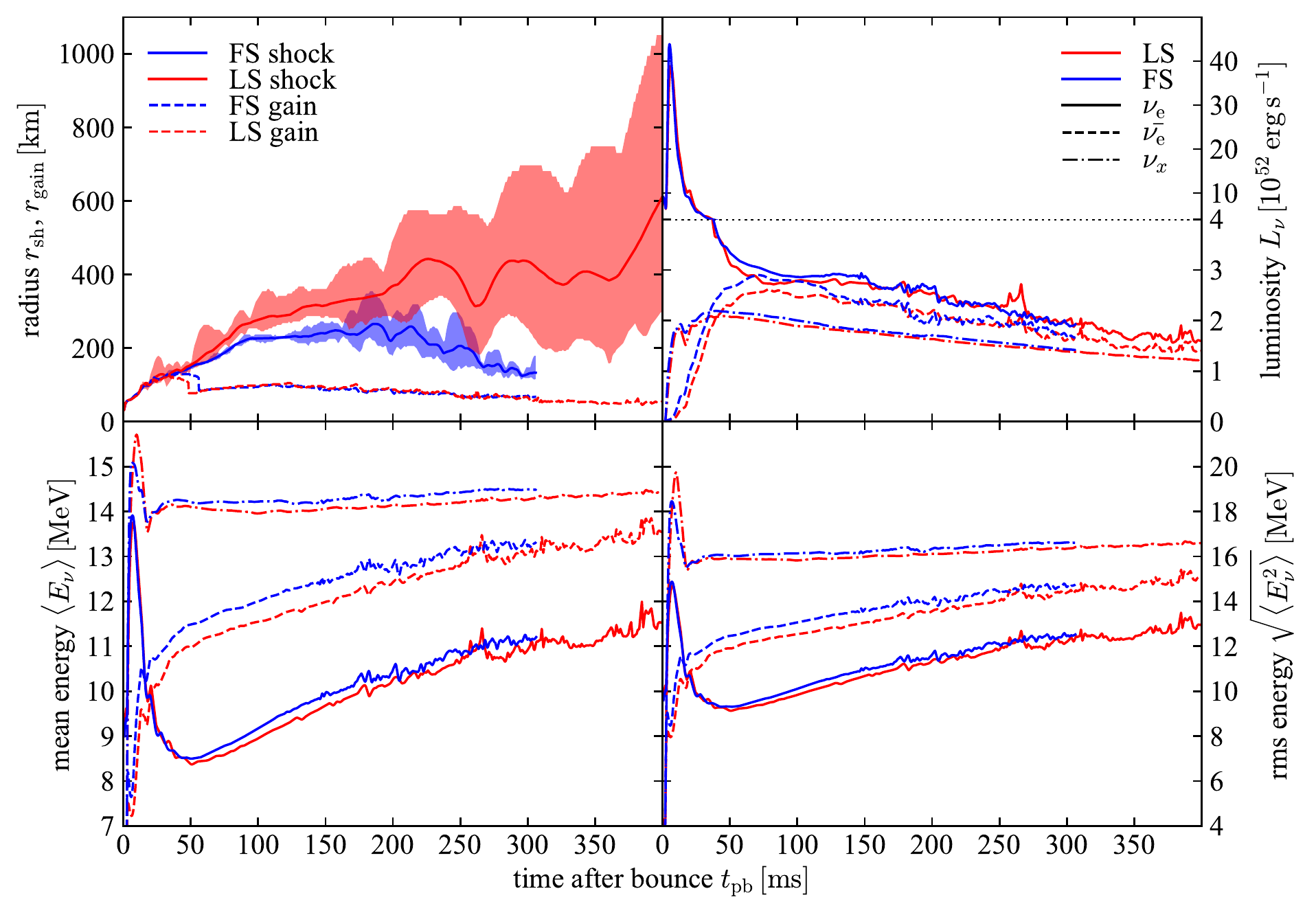}
\caption{The time evolution of several key quantities. For each panel, red and blue represent the LS and the FS models, respectively. The upper-left panel shows the shock (solid lines and filled regions) and the gain (dashed lines) radii. The solid lines are the mean shock radii, and the upper and lower edges of the filled regions indicate the maximum and the minimum shock radii. The upper-right, lower-left, and lower-right panels represent the neutrino luminosities, the mean energies, and the rms energies for $\nue$ (solid line), $\nueb$ (dashed line), and $\nux$ (dash--dotted line), respectively. Both luminosities and the energies are measured at $r=500\,{\rm km}$. The upper-right panel is divided into two parts, in order to show the neutronization burst for $\nue$ and the later light curve at once. The upper and lower halves have different vertical scales each other. \label{fig:evolutions}}
\end{figure*}

First of all, we present several quantities to summarize the supernova dynamics in figure \ref{fig:evolutions}: the shock evolutions, the neutrino luminosities, the neutrino mean energies, and the root-mean-square (rms) energies. Throughout this paper, the shock radius is defined by the outermost radius where the radial velocity gradient becomes negative. The maximum shock radius reaches $\sim 1000\,{\rm km}$ for the LS model. The FS model shows a contracting mean shock radius despite the neutrino luminosities and mean energies for the FS model being slightly higher than those for the LS model. Because the maximum and mean shock radii show continuous expansion, we regard the LS model as a successful shock revival model.

\begin{figure*}[ht!]
\plotone{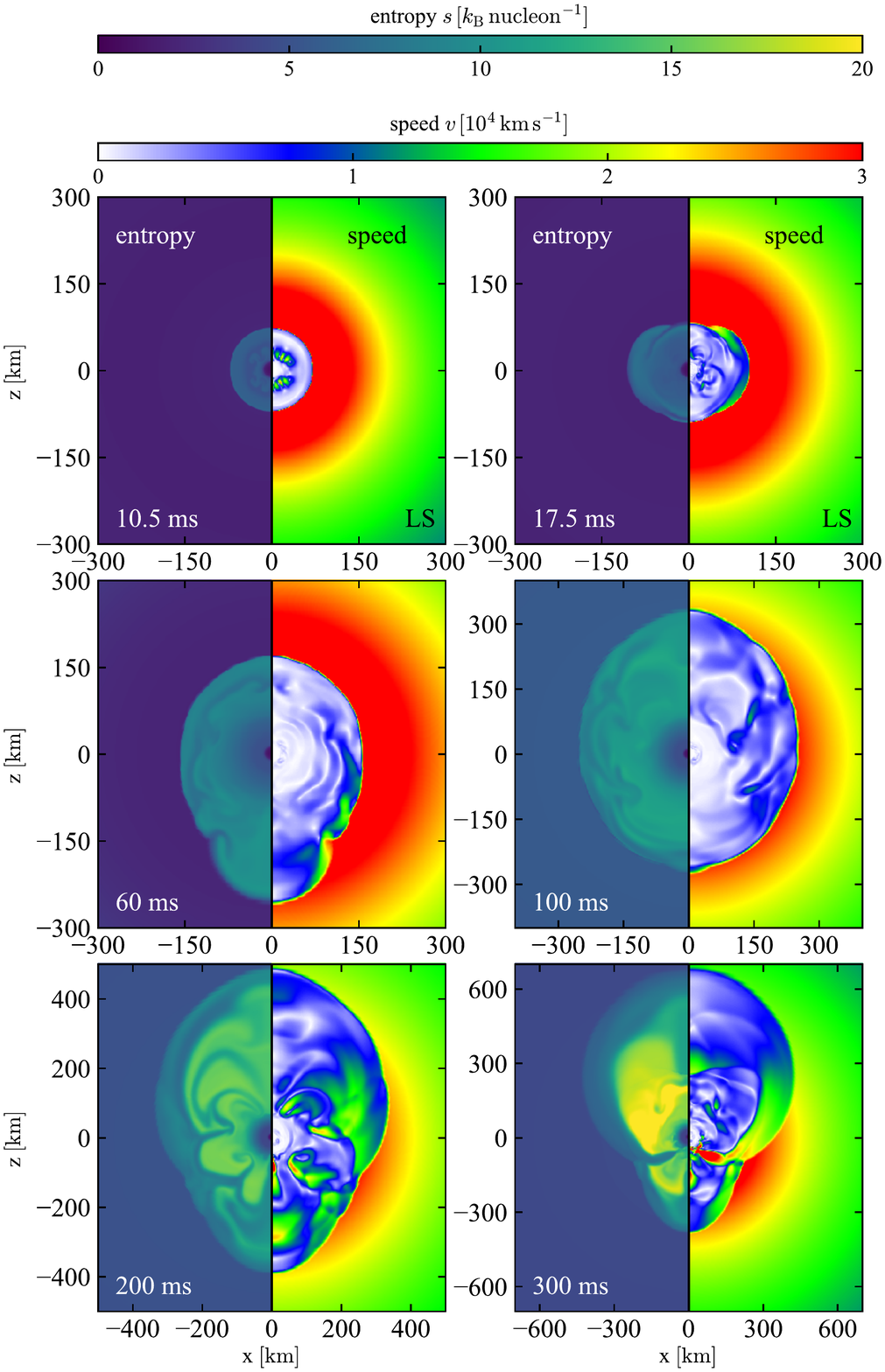}
\caption{Snapshots of the entropy and the velocity of the LS model. For each panel, the left half shows the entropy distribution and the right half displays the speed distribution. The time after the bounce is indicated at the bottom left of each figure. \label{fig:LSsnapshots}}
\end{figure*}

\begin{figure*}[ht!]
\plotone{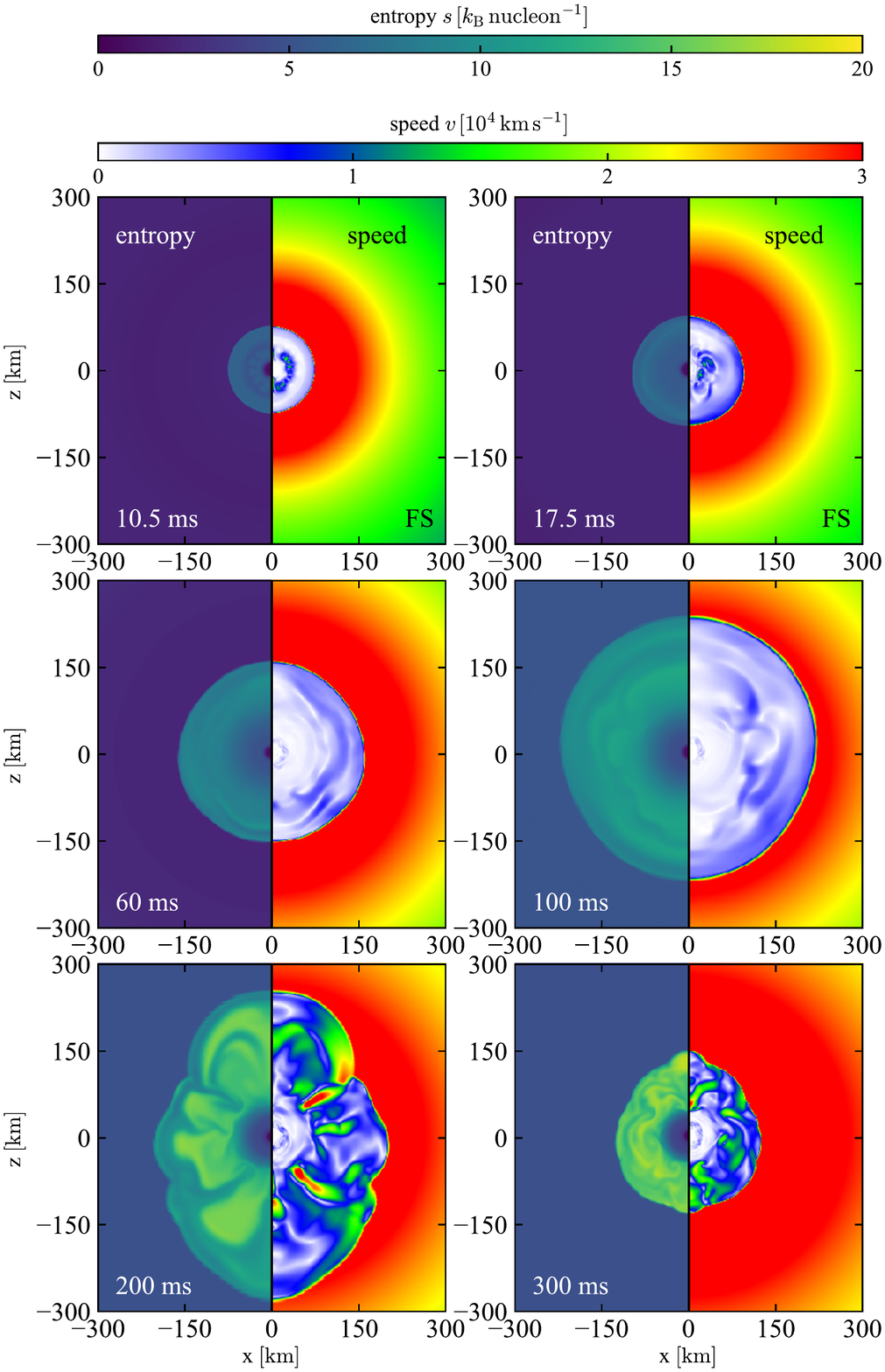}
\caption{The same as figure \ref{fig:LSsnapshots} except that the snapshots of the FS model are shown. \label{fig:FSsnapshots}}
\end{figure*}

To illustrate the dynamical features of our simulations, we show the snapshots of the entropy and the speed $v:=\sqrt{(v^r)^2 + (rv^\theta)^2}$ at different times for the LS and FS models in figures \ref{fig:LSsnapshots} and \ref{fig:FSsnapshots}, respectively. Note that $v^\theta$ is the angular velocity, which is usually measured in units of ${\rm rad\,s^{-1}}$. At the onset of the prompt convection ($t_{\rm pb}=10.5\,{\rm ms}$), a slightly stronger convective motion is seen in the LS model than in the FS model. As a consequence, the LS model shows a more aspherical, violent shock deformation at $t_{\rm pb}=17.5\,{\rm ms}$, while the sizes of the shock themselves are similar for both models. Then, the shock radii of both models expand gradually as seen in the snapshots at $t_{\rm pb}=60\,{\rm ms}$ and $t_{\rm pb}=100\,{\rm ms}$. Again, the shape of the shock in the LS model is more aspherical than that in the FS model. At $t_{\rm pb}=200\,{\rm ms}$ and $t_{\rm pb}=300\,{\rm ms}$, the LS model indicates continuous shock expansion. Although the FS model shows vigorous shock deformation at $t_{\rm pb}=200\,{\rm ms}$, the shock contracts after that, as seen in the snapshot at $t_{\rm pb}=300\,{\rm ms}$.

\begin{figure*}[ht!]
\plottwo{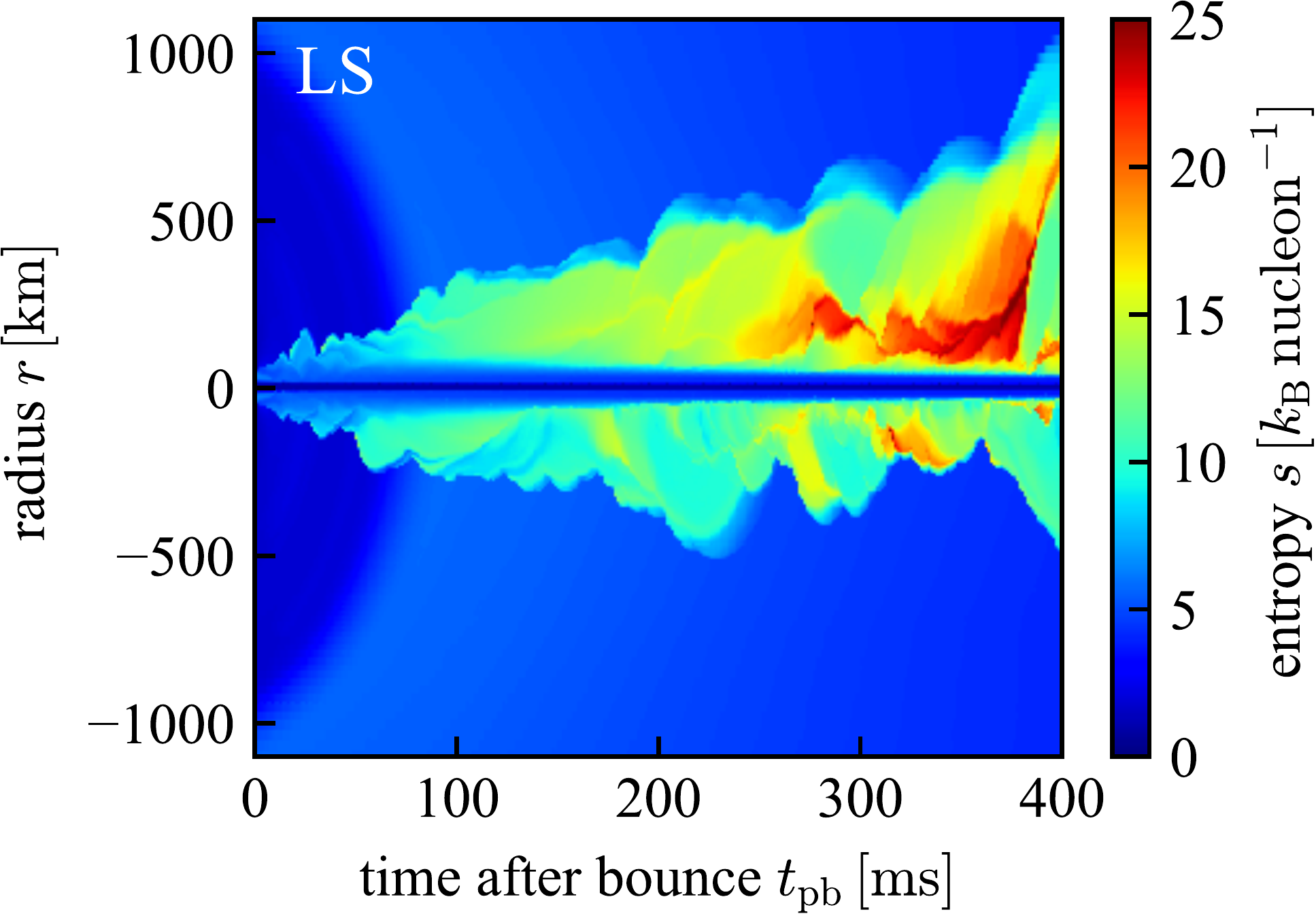}{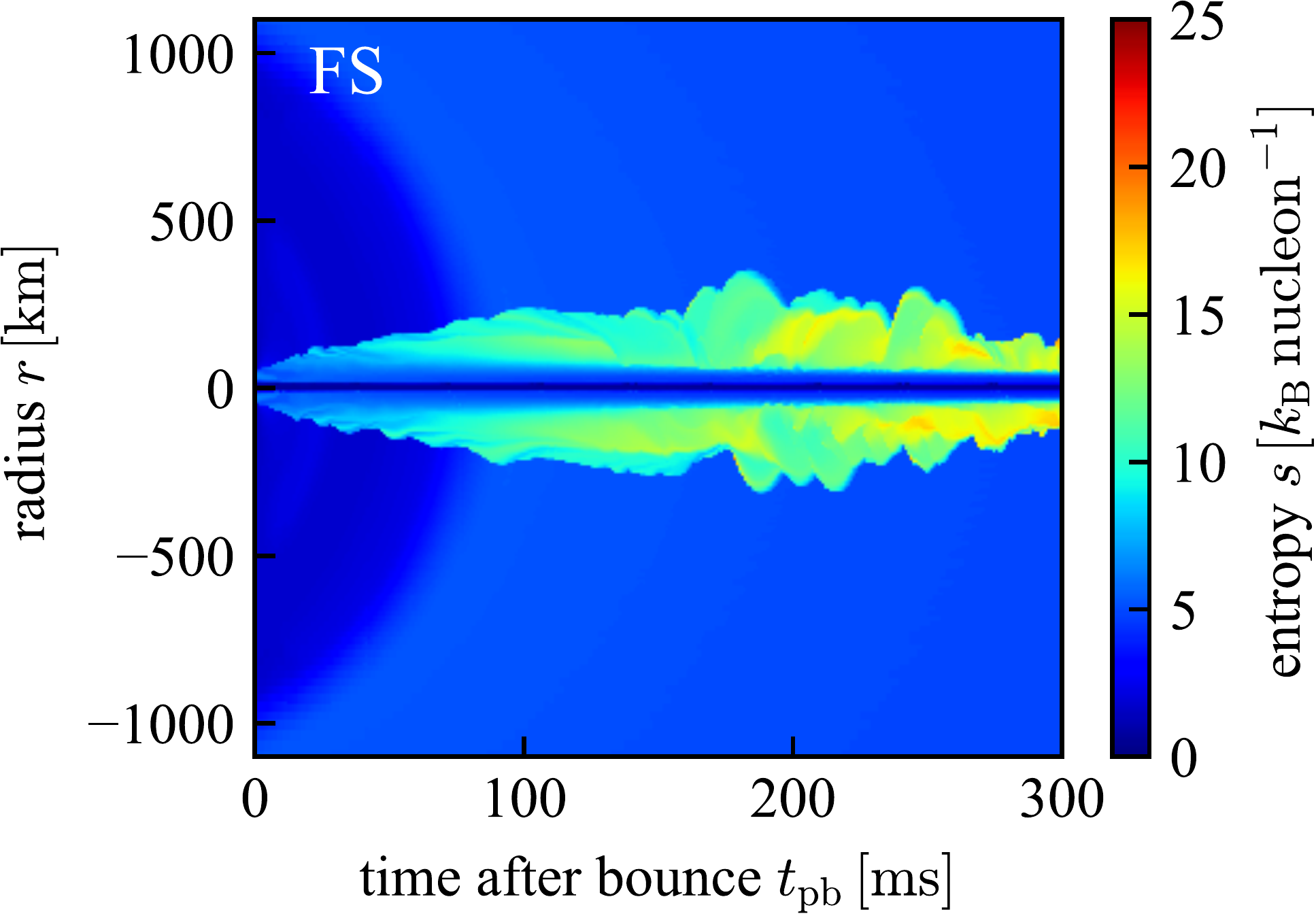}
\caption{The time evolution of the entropy profiles along the north and south poles. The left panel shows the LS model, while the right panel represents the FS model. The horizontal scale is different between the two panels because the simulations are stopped at different times: $400\,{\rm ms}$ for the LS model and $300\,{\rm ms}$ for the FS model. \label{fig:entropy_NS}}
\end{figure*}

In figure \ref{fig:entropy_NS}, we show the time evolution of the entropy along the north ($\theta=0$) and the south ($\theta=\pi$) poles for both models up to the end of the simulations. Again, we can see that the LS model shows gradual but continual shock expansion, especially along the north pole, while the FS model does not show shock expansion.

\begin{figure}[ht!]
\plotone{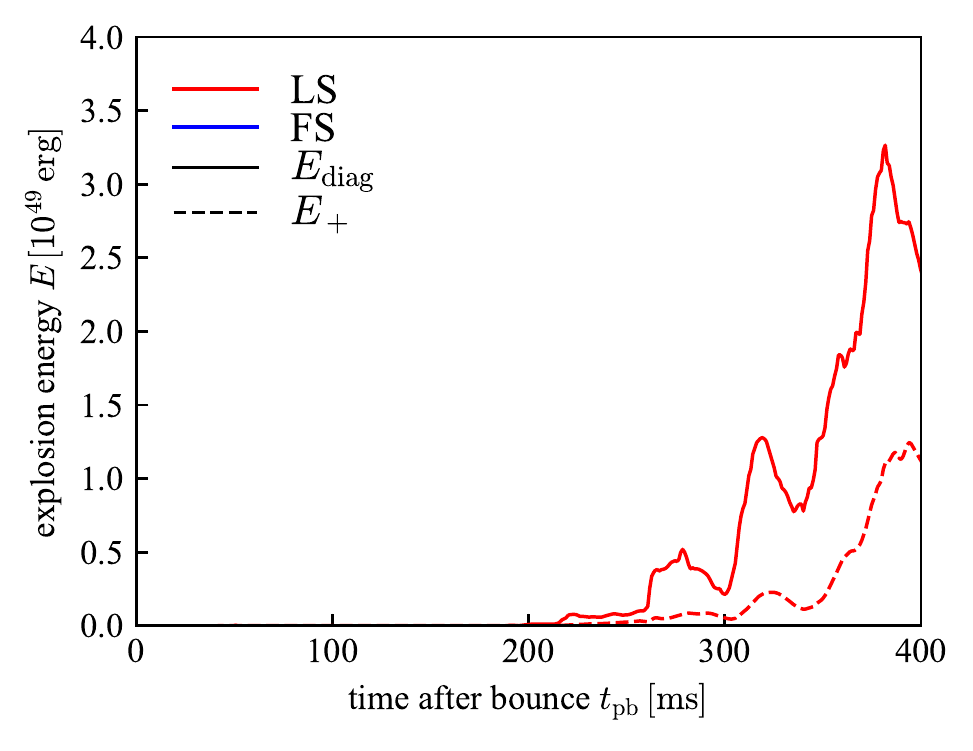}
\caption{The explosion energies of the LS (red) and FS (blue) models. The solid and dashed lines represent the total positive energy $E_+$ and the diagnostic explosion energy $E_{\rm diag}$, respectively. Although the legend for the FS model is indicated, the explosion energy for the FS model is zero. \label{fig:diagnostic_energy}}
\end{figure}

Figure \ref{fig:diagnostic_energy} shows the explosion energies of the LS and FS models. The total energy density is defined by
\begin{equation}
e_{\rm tot} := \frac{1}{2}\rho v^2 + e_{\rm th} + \rho \psi \label{eq:totene}
\end{equation}
in this paper. Here,
\begin{equation}
e_{\rm th} := \rho \sum_{i\in{\rm nuclear\,species}} \frac{3}{2}\frac{k_{\rm B}T}{m_{\rm u}}\frac{X_i}{A_i} + aT^4 + \left(e_{\rm e^-e^+}-\rho Y_{\rm e}\frac{m_{\rm e}c^2}{m_{\rm u}}\right) \label{eq:thermal}
\end{equation}
is the thermal energy defined in \citet{2016ApJ...818..123B}; $X_i$, $A_i$, $a$, $e_{\rm e^-e^+}$, and $m_{\rm e}$ are the mass fraction and the mass number of nuclear species $i$, the radiation constant, the internal energy density for electrons and positrons including rest-mass energy, and the mass of the electron, respectively. For the moment, we explicitly write the light speed $c$ to emphasize that the rest-mass energy of the electron is subtracted. In figure \ref{fig:diagnostic_energy}, we consider two kinds of explosion energies: the total positive energy $E_+$ and the diagnostic explosion energy $E_{\rm diag}$. The total positive energy is the total energy integrated over the region where the total energy is positive:
\begin{equation}
E_+ := \int_{e_{\rm tot}>0} e_{\rm tot} \rd V.
\end{equation}
The diagnostic explosion energy is similar to the total positive energy, but the integrand contains the nuclear binding energy:
\begin{equation}
E_{\rm diag} := \int_{e_{\rm tot}>0} (e_{\rm tot} + [e_{\rm bind}({}^{56}{\rm Fe}) - e_{\rm bind}])  \rd V.
\end{equation}
The binding energy of iron $e_{\rm bind}({}^{56}{\rm Fe})=8.8\,{\rm MeV}\times \rho/m_{\rm u}$ is the nuclear binding energy if the all nucleons were used up to form ${}^{56}{\rm Fe}$. The binding energy $e_{\rm bind}$ is the actual nuclear binding energy evaluated by the internal energy density subtracted by the thermal energy $e_{\rm th}$.

Although the diagnostic explosion energy for the LS model is much smaller than $10^{51}\,{\rm erg}$ at the end of the simulation, it seems to grow at the rate of $\sim 2\times 10^{50}\,{\rm erg\,s^{-1}}$. On the other hand, the explosion energies of the FS model is zero as this model fails to explode.

\begin{figure}[ht!]
\plotone{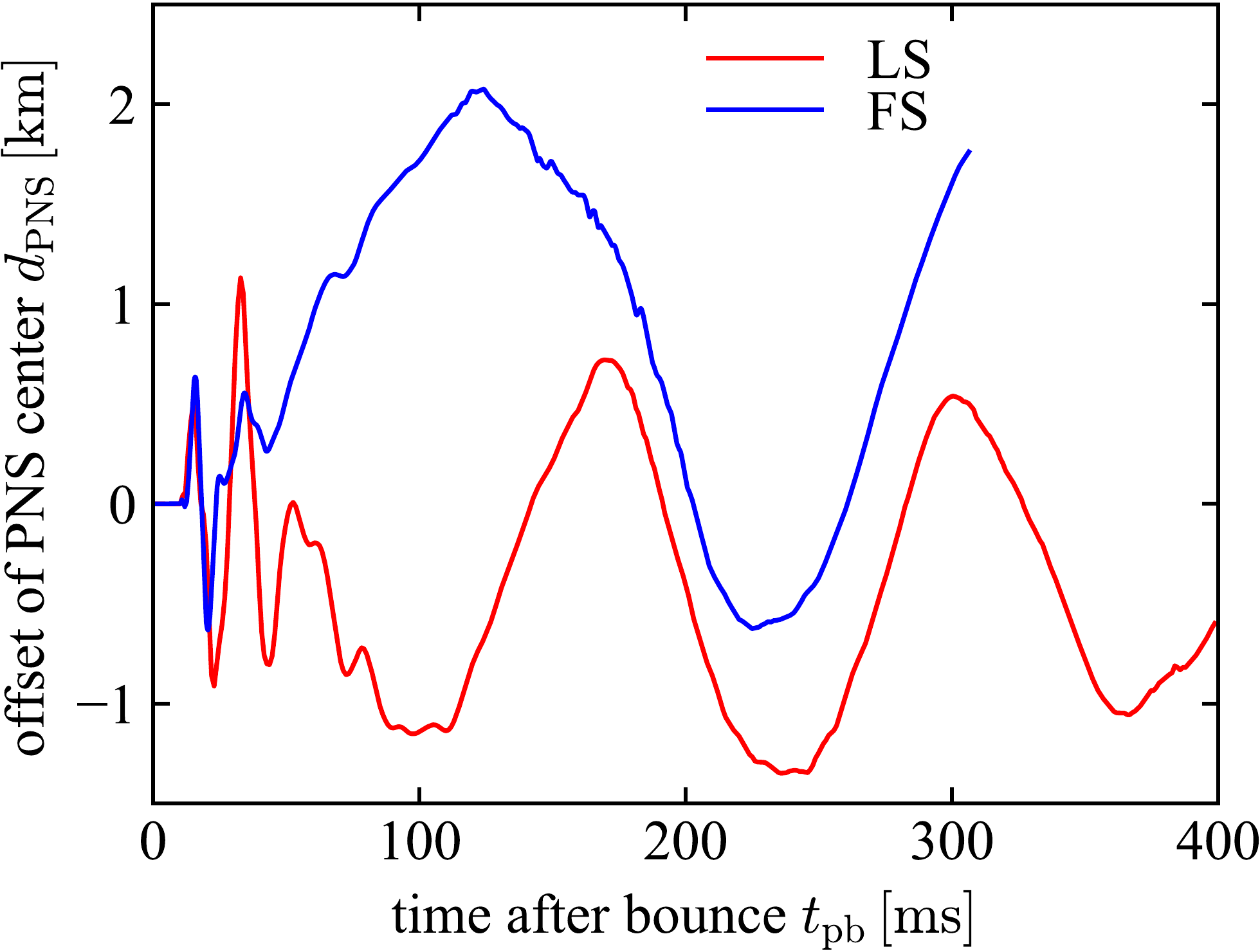}
\caption{The trajectory of the PNS center along the symmetry axis. The red and blue lines are for the LS and FS models, respectively. \label{fig:PNSoffset}}
\end{figure}

The central PNSs move due to the recoil of the asymmetric expansion of the matter, but the kick velocities are small. Figure \ref{fig:PNSoffset} indicates the trajectories of the PNS centers along the symmetry axis for both EOS models. Tracking the trajectory of the PNS center is one of the unique features of our code because we utilize the acceleration frame. As shown in the figure, the offsets are a few kilometers and the velocities are $\sim 10\,{\rm km\,s^{-1}}$. They are not very violent; thus, the differences in the acceleration and the laboratory frames for both models are small. Therefore, we ignore the difference between these frames in this paper. The larger kick is possibly seen after the time we stopped the simulation, or with simulations with other settings such as the progenitor, the rotation, the EOSs, and so on. An example is discussed in \citet{2019ApJ...880L..28N}.\footnote{\citet{2019ApJ...880L..28N} employ a different momentum feedback scheme from that in this paper. The new and more accurate feedback scheme is described in \citet{2019ApJ...878..160N}. The accuracy of the feedback scheme is not crucial for the analysis in this paper because the same treatment is employed for both models, though the kick velocities themselves are possibly not robust.}

\subsection{Deformation of the shock with the convection} \label{sec:defshock}

\begin{figure*}[ht!]
\plottwo{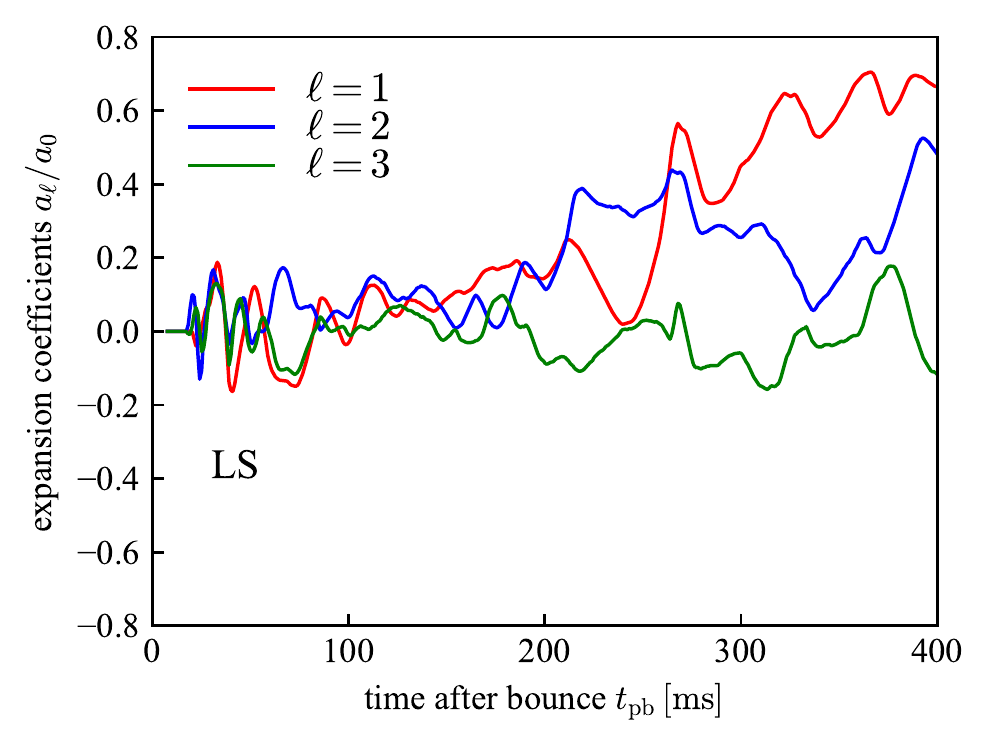}{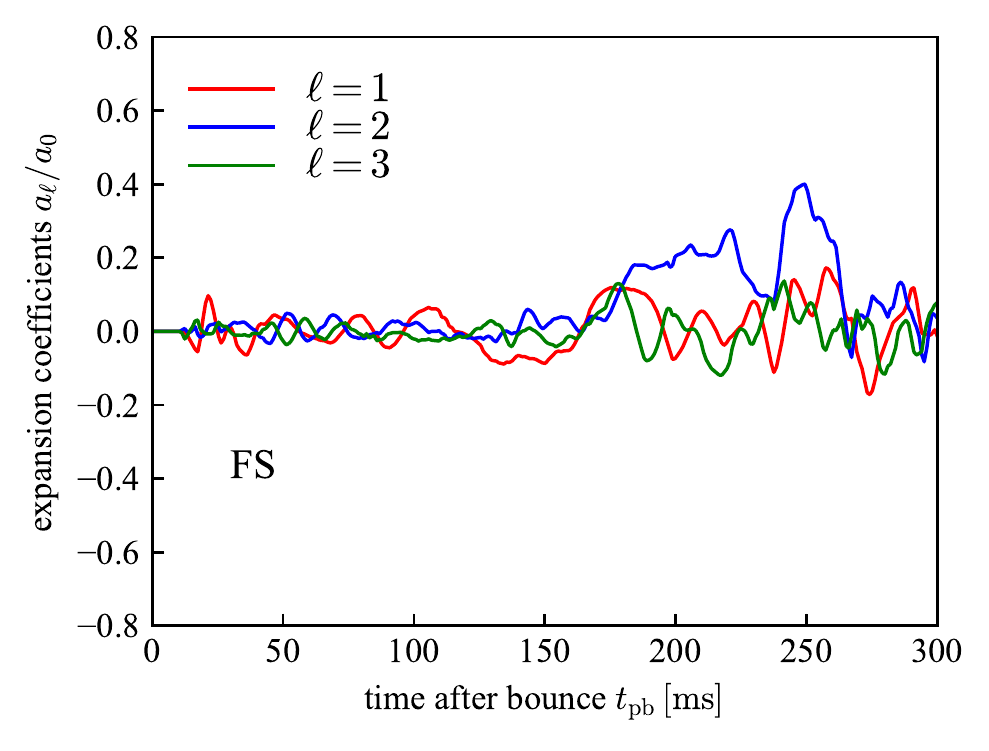}
\caption{The Legendre expansion coefficients of the shock radius. The left and right panels indicate the coefficients for the LS and FS models, respectively. The different colors correspond to different modes: red, blue, and green for the $\ell=1$, $2$, and $3$ modes, respectively. Each coefficient is normalized by the mean shock radius, i.e., the $\ell=0$ mode coefficient $a_0$. \label{fig:shockmode}}
\end{figure*}

The snapshot at $t_{\rm pb} = 17.5\,{\rm ms}$ for the LS model (the top-right panel of figure \ref{fig:LSsnapshots}) indicates that the shock in the LS model deforms significantly. To compare the deformation of the shock, we decompose the shock shape into Legendre polynomials. Figure \ref{fig:shockmode} indicates the time evolution of the expansion coefficients normalized by the mean shock radii. The expansion coefficients are defined by
\begin{equation}
a_\ell := \frac{2\ell + 1}{2}\int_{0}^\pi r_{\rm sh}(\theta) P_\ell(\cos\theta) \sin\theta \rd \theta, \label{eq:coef}
\end{equation}
where $r_{\rm sh}(\theta)$ and $P_\ell(\cos \theta)$ are the shock radius as a function of $\theta$ and Legendre polynomial of degree $\ell$, respectively. The mean shock radius is nothing but $a_0$. Here, we show $\ell=1$, $2$, and $3$ modes in the figure. The mode amplitudes are much larger for the LS model than the FS model.

Although oscillations of the shock shape in the large-scale modes at the early postbounce phase are seen in figure \ref{fig:shockmode}, this is not evidence of the strong standing accretion shock instability (SASI) motion. The SASI is the instability of the standing shock by definition. Because the top-right panel of figure \ref{fig:LSsnapshots} is in the prompt convection phase with an initially expanding shock, this is not the SASI. This deformation of the shock arises from the interaction with the accretion flow outside the shock and the convection inside the shock. When the rising convective parcels reach the shock surface, they are refracted along the shock. The newly accreting flow is involved with this refracting flow, and then it overrides the neighboring shock front. This motion produces violent deformation in the top-right panel of figure \ref{fig:LSsnapshots}.

\begin{figure}[ht!]
\plotone{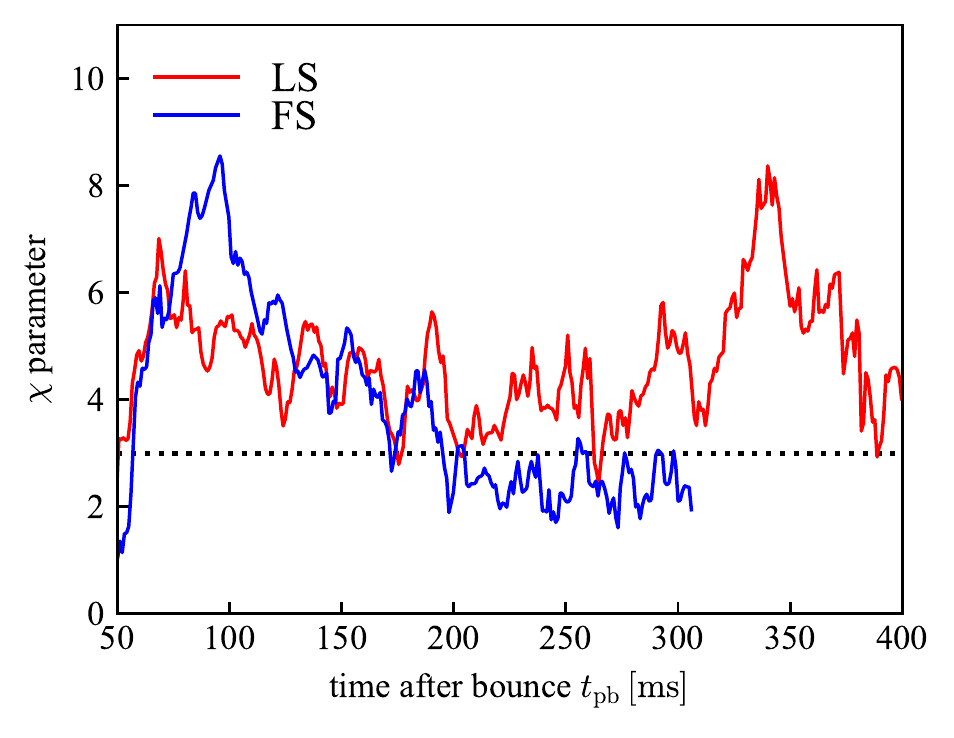}
\caption{The time evolution of Foglizzo's $\chi$ parameter. The red and blue lines represent the LS and FS models, respectively. The horizontal dotted line indicates $\chi=3$, where the class of the fluid instability bifurcates. \label{fig:chiparameter}}
\end{figure}

The deformation of the shock at the standing shock phase, around $t_{\rm pb} \sim 100\,{\rm ms}$, originates from the neutrino-driven convection. Figure \ref{fig:chiparameter} shows the Foglizzo's $\chi$ parameter \citep{2006ApJ...652.1436F}; neutrino-driven convection develops when $\chi \ga 3$, while the SASI develops when $\chi \la 3$. Here, $\chi$ parameter is defined by
\begin{equation}
\chi = \int_{\langle r_{\rm gain} \rangle}^{\langle r_{\rm sh} \rangle} \frac{\omega_{\rm BV,1D}}{\langle v^r \rangle} \rd r, \label{eq:chi}
\end{equation}
where
\begin{equation}\omega_{\rm BV,1D} = \sqrt{\Big| \left(\frac{1}{\langle \gamma \rangle \langle p \rangle}\frac{\partial \langle p \rangle}{\partial r} - \frac{1}{\langle \rho \rangle}\frac{\partial \langle \rho \rangle}{\partial r} \right)\frac{\partial \psi}{\partial r}\Big|}
\end{equation}
is the Brunt--V\"ais\"al\"a frequency, and $\gamma$ is the adiabatic index. The angle brackets represent the angular average of a physical quantity: $\langle \bullet \rangle := \int \bullet \rd \Omega /4\pi$. The original definition in \cite{2006ApJ...652.1436F} is meant to be applied under spherical symmetry, and hence equation (\ref{eq:chi}) is evaluated from the angular-averaged profiles similarly to \citet{2014MNRAS.440.2763F, 2014ApJ...786..118I}. Note that the $\chi$ parameter in figure 1 of \citet{2018ApJ...854..136N} is defined by the integral from the average gain radius to the maximum shock radius, and hence it is slightly different from the $\chi$ parameter in this paper. In figure \ref{fig:chiparameter}, we consider the time after $t_{\rm pb}=50\,{\rm ms}$ because the gain region appears only after that time. Figure \ref{fig:chiparameter} clearly shows that $\chi > 3$ for both LS and FS models before $t_{\rm pb} \sim 200\,{\rm ms}$, and hence, neutrino-driven convection develops first in both models. After $t_{\rm pb} \sim 200\,{\rm ms}$, the $\chi$ parameter in the FS model becomes smaller than 3. Hence, the SASI may develop in the FS model, but the highly developed turbulence makes it difficult to distinguish the class of the fluid instability, the convection or SASI, at that time.

Compared to \citet{2012ApJ...756...84M, 2012ApJ...749...98T, 2014ApJ...786...83T} , the $\ell=1$ mode amplitudes of the shock in figure \ref{fig:shockmode} are $\sim 10\text{--} 10^2$ times larger. We impose the initial perturbations on our simulations in a different way from their simulations. The radial velocity perturbation with an amplitude of $0.1\%$ is imposed in the preshock region at $t_{\rm pb} \sim 0.6\,{\rm ms}$ in our simulations, while the velocity perturbation with an amplitude of $1\%$ is employed in the postshock region at $t_{\rm pb} = 10\,{\rm ms}$ in \citet{2014ApJ...786...83T}, for example. According to \citet{2016MNRAS.461.3864A}, the energy of the preshock perturbation (squared velocity) is amplified by a factor of $\sim 2$ when passing the shock front, and it plays a minor role here. Owing to the prompt convection, the perturbation initially given at $\mathcal{O}(0.1)\,{\rm ms}$ grows much larger than $10 \times$ at $t_{\rm pb}=10\,{\rm ms}$ as estimated in the following. The convective growth rate, or Brunt--V\"ais\"al\"a frequency, is approximately \citep[see, c.f., figure 2 in][]{2018ApJ...854..136N}
\begin{eqnarray}
\omega_{\rm BV} &\sim& \sqrt{\frac{GM_{\rm PNS}}{r^2}\frac{\pa s/k_{\rm B}}{\pa r}} \nonumber \\ 
&=& 2.4\,{\rm ms^{-1}} \nonumber \\
&\times&
\left(\frac{M_{\rm PNS}}{0.7\,M_\odot}\right)^{1/2}\left(\frac{r}{40\,{\rm km}}\right)^{-1}\left( \frac{\pa (s/k_{\rm B})/\pa r}{10^{-6}\,{\rm cm^{-1}}} \right)^{1/2}, \label{eq:BVfreq}
\end{eqnarray}
where the gravitational acceleration is approximated by
\begin{equation}
\frac{\pa \psi}{\pa r} \sim \frac{GM_{\rm PNS}}{r^2},
\end{equation}
and $M_{\rm PNS}$ is the PNS mass. As long as the perturbation is linear, the initial perturbation at $\mathcal{O}(0.1)\,{\rm ms}$ grows $\exp(10\,{\rm ms} \times \omega_{\rm BV})\sim 3\times 10^{10}$ times larger at $10\,{\rm ms}$. 
Such a large amplification breaks the linear approximation, and the actual amplification should be smaller than this estimation. It is still certain that the initial $0.1\%$ velocity perturbation at $0.6\,{\rm ms}$ like ours grows to exceed $1\%$ velocity perturbation in \citet{2014ApJ...786...83T} at $10\,{\rm ms}$. Because the initial perturbation is effectively large in this work, the prompt convection and hence the shock deformation are much larger than other works. Although the central iron core is convectively stable, a possible slight perturbation imparted by the convective silicon layer can be amplified during the core collapse \citep{2000ApJ...540..962H}. Hence, the early initial perturbation employed in this paper is possible.

\subsection{Timescale ratio and the effects of the nuclear composition of the EOSs} \label{sec:timescale}

Next, we evaluate the timescale ratio $\tau_{\rm adv}/\tau_{\rm heat}$ to see how close our two simulations are to success. If this ratio exceeds unity, successful shock revival occurs. Here, $\tau_{\rm adv}:=M_{\rm gain}/\dot{M}_{\rm sh}$ and $\tau_{\rm heat} := |E_{\rm gain}|/Q_{\rm gain}$ are the advection timescale through the gain layer and the heating timescale, respectively, where $M_{\rm gain}$, $\dot{M}_{\rm sh}$, $E_{\rm gain}$, and $Q_{\rm gain}$ are the gain mass, the mass accretion rate at the shock, the total energy in the gain layer, and the net heating rate in the gain layer, respectively. The mass accretion rate is defined as $\dot{M}_{\rm sh}:=-\int_{\rm shock}4\pi \rho v^r r^2 \rd \Omega$. In this paper, we define $E_{\rm gain}$ as
\begin{equation}
E_{\rm gain} := \int_{\rm gain}e_{\rm tot}\rd V,
\end{equation}
where $e_{\rm tot}$ is defined in equation (\ref{eq:totene}).

The definition of the timescale ratio in this paper is improved from that in the previous paper \citep{2018ApJ...854..136N}. They are different in two respects: the definition of the total energy and the domain of the integral are different. In the previous paper, the total energy includes the nuclear binding energy, whose offset we are free to choose. The total energy defined by equation (\ref{eq:totene}) with the thermal energy defined by equation (\ref{eq:thermal}) is free from such ambiguity. The integral in the previous paper is the radial integral of the angular-averaged quantities from the mean gain radius to the minimum shock radius. On the other hand, the angular dependence of the gain radii and shock radii is taken into account in this paper.

\begin{figure*}[ht!]
\plotone{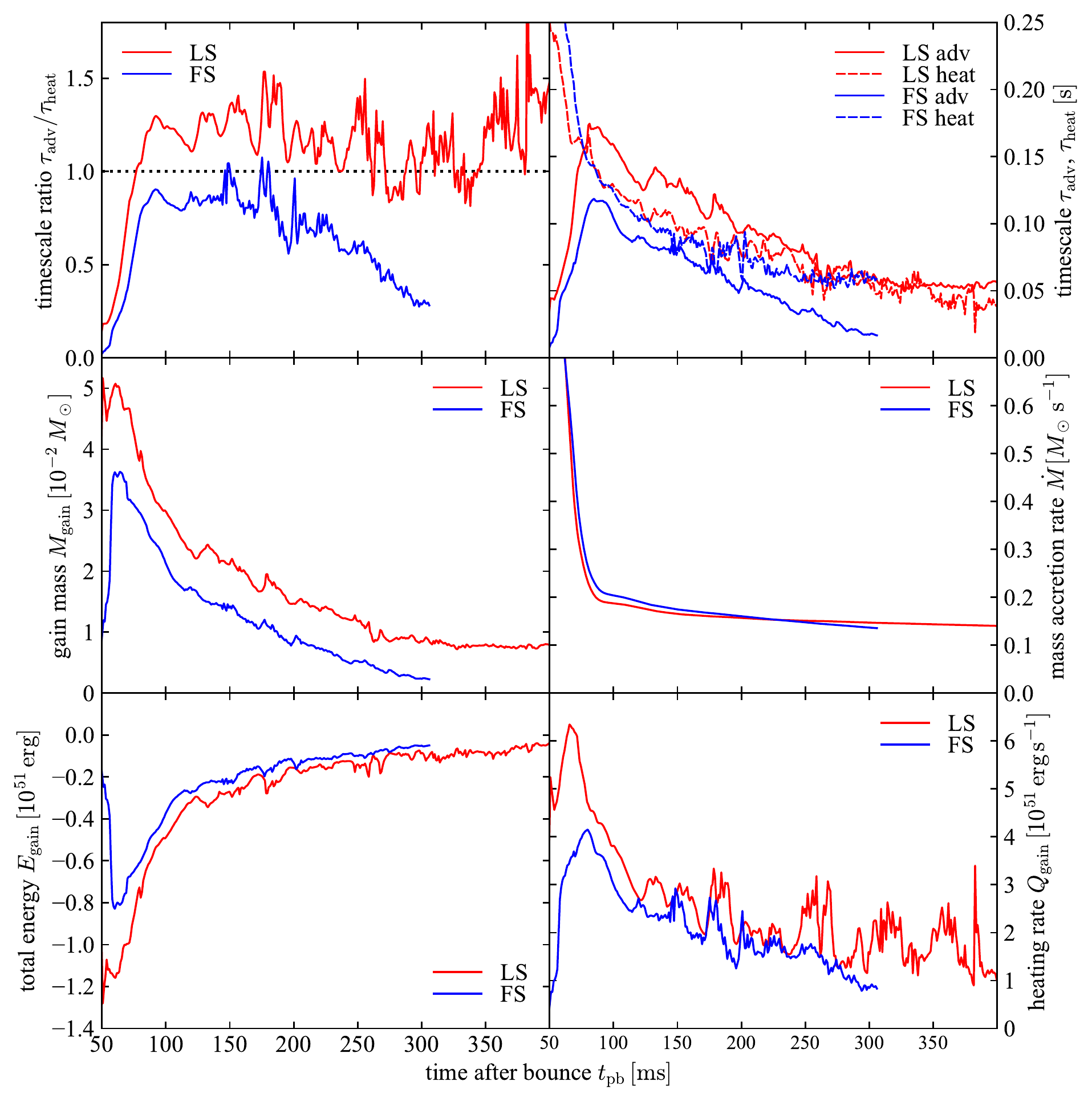}
\caption{The evolution of the timescale ratio $\tau_{\rm adv}/\tau_{\rm heat}$ and related quantities. For all panels, the red and blue lines correspond to the LS and FS models, respectively. The top-left panel indicates the timescale ratio $\tau_{\rm adv}/\tau_{\rm heat}$ with the horizontal dotted line indicating unity. The top-right panel shows the advection timescale (solid lines) and the heating timescale (dashed lines) themselves. The middle-left, middle-right, bottom-left, and bottom-right panels display the gain mass, the mass accretion rate, the total energy in the gain region, and the net heating rate in the gain region, respectively. \label{fig:timescale}}
\end{figure*}

The timescale ratio in the LS model exceeds unity for most of the time after $\sim 70\,{\rm ms}$ as indicated in figure \ref{fig:timescale}. On the other hand, the FS model shows a ratio smaller than unity for almost the entirety of the simulation period. This implies that the LS model is favorable for a successful explosion. It is worth noting that the timescale ratio of the FS model exceeds unity in \citet{2018ApJ...854..136N} due to the inappropriate treatment of the thermal energy. Thanks to equation (\ref{eq:thermal}), more reliable evaluations of the timescale ratio are obtained.

A closer look reveals to us that the heating timescales are similar for both models, while the advection timescales are much larger for the LS model. For the advection timescales, the mass accretion rate is not that different between the two models, but the gain mass is much larger for the LS model. For the heating timescales, the total energy in the gain region $E_{\rm gain}$ is slightly lower for the LS model, but this is compensated by the slightly higher net heating rate $Q_{\rm gain}$.

The reason for the difference in the gain mass is the strength of the turbulent motion. The right-side parts of each panel of figures \ref{fig:LSsnapshots} and \ref{fig:FSsnapshots} show the speed of matter $v = \sqrt{(v^r)^2 + (rv^\theta)^2}$. From the snapshots at $t_{\rm pb}=10.5\,{\rm ms}$, the speed of matter at the prompt convection in the LS model is more vigorous than that in the FS model. This stronger prompt convection in the LS model gives stronger seed perturbations of the later neutrinodriven convection. This develops stronger turbulence. This mechanism was also suggested in the previous paper \citep{2018ApJ...854..136N}, but we present a more detailed description here.

\begin{figure}[ht!]
\plotone{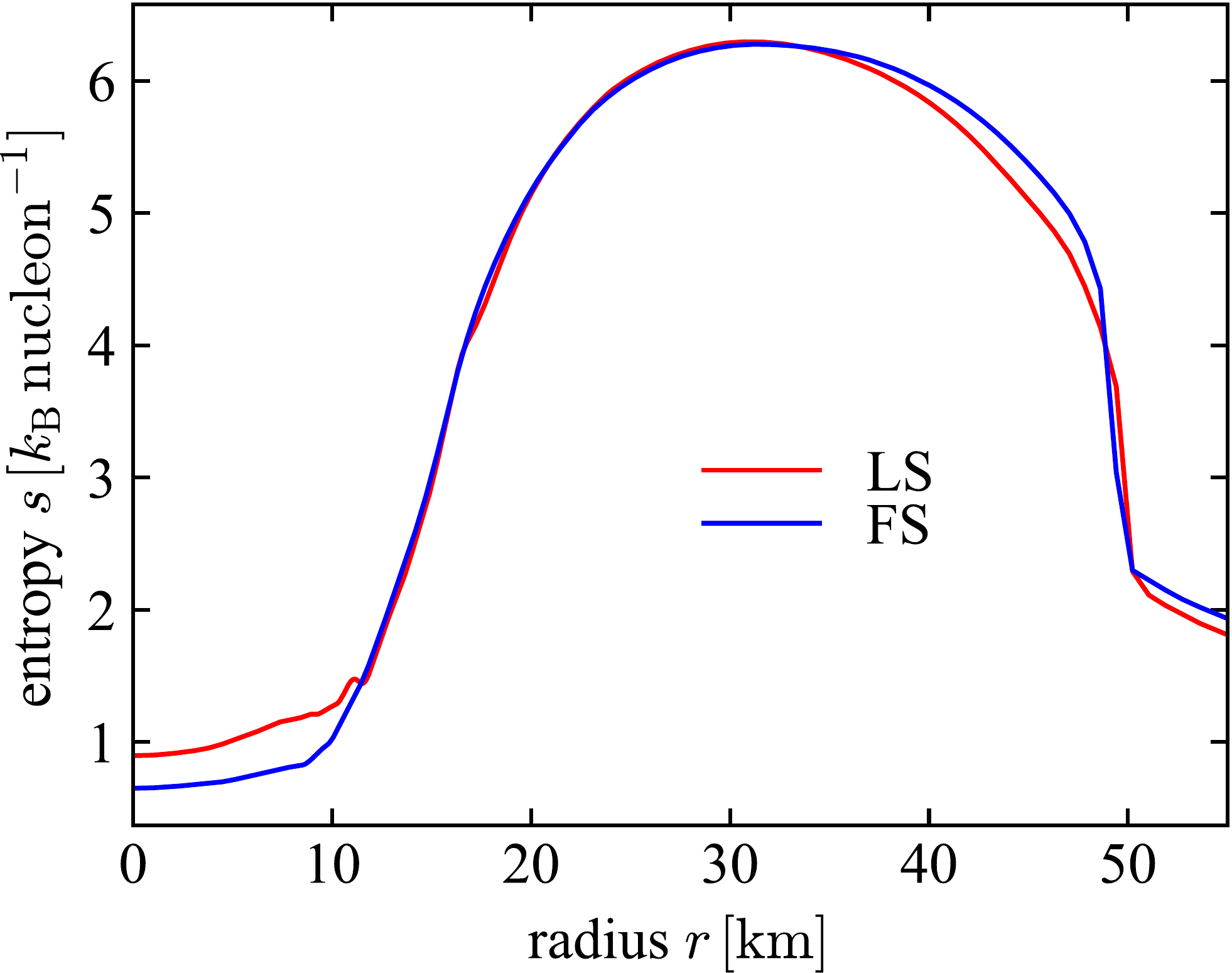}
\caption{The radial entropy profiles at $t_{\rm pb} \simeq 1.7\,{\rm ms}$, when the shock waves have swept the convectively unstable region. The red and blue lines are for the LS and FS models, respectively. \label{fig:LSFSentrograd_prompt}}
\end{figure}

The origin of the stronger prompt convection is the difference in the radial gradient of the entropy profiles. The convectively unstable regions at $t_{\rm pb} = 10.5\,{\rm ms}$ in figures \ref{fig:LSsnapshots} and \ref{fig:FSsnapshots} are up to a few tens of kilometers. Figure \ref{fig:LSFSentrograd_prompt} shows the entropy profiles at the time when the shock waves have swept the convectively unstable region. Due to the weakening of the propagating shock, the radial gradients of the entropy are negative in $35\,{\rm km} \la r \la 50\,{\rm km}$ for both models. The gradient is steeper for the LS model than for the FS model. Note that the value of the entropy gradient discussed in section \ref{sec:defshock} is estimated from this figure.

\begin{figure}[ht!]
\plotone{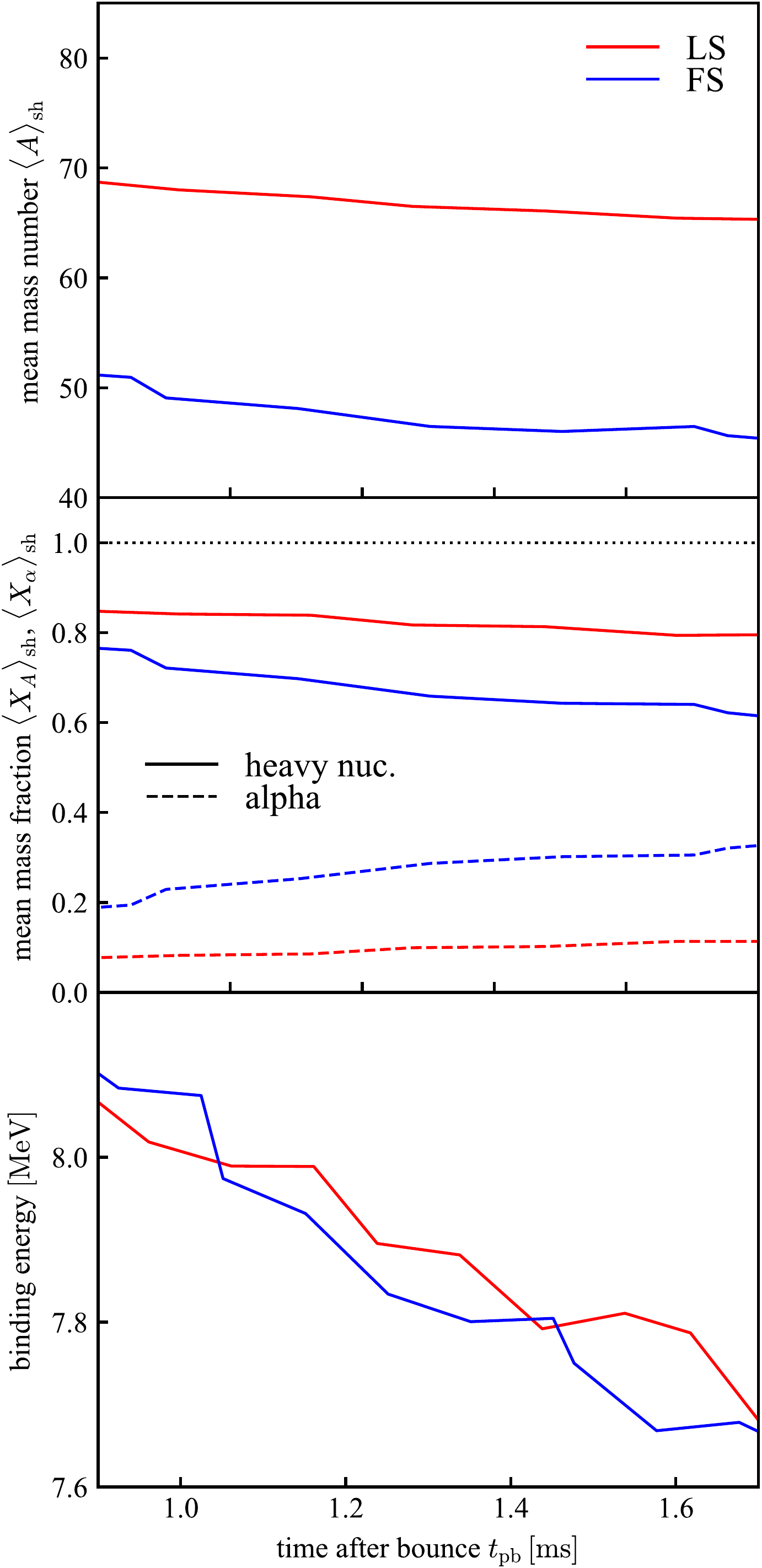}
\caption{The nuclear composition of the accreting matter just above the shock for the LS and FS models during the period relevant to the growth rate of the prompt convection. The top panel shows the mass number $\langle A \rangle_{\rm sh}$. The middle panel displays the mass fraction $\langle X_{A,\,\alpha}\rangle_{\rm sh}$ of the representative heavy nucleus (solid lines) and the alpha particle (dashed lines). The horizontal dotted line indicates unity. The bottom panel indicates the estimated nuclear binding energy per nucleon of the accreting matter. For all panels, the red and blue lines correspond to the LS and FS models, respectively. \label{fig:composition}}
\end{figure}

Because an important player in the shock weakening is the photodissociation of heavy nuclei, we compare the nuclear composition between the LS and FS models in figure \ref{fig:composition}. The time when the shock sweeps the region $35\,{\rm km} \la r \la 50\,{\rm km}$ is merely $0.9\,{\rm ms} \la t_{\rm pb} \la 1.7\,{\rm ms}$. Figure \ref{fig:composition} shows the average mass number $\langle A\rangle_{\rm sh}$ and the average mass fraction $\langle X_A\rangle_{\rm sh}$ of the matter slightly outside the shock during this period for both the LS and FS models. Also, the mass fraction of alpha particle $\langle X_\alpha\rangle_{\rm sh}$ is shown. The angular-averaged quantity of the accreting matter $\langle \bullet \rangle_{\rm sh}$ is defined as $\int_{r_{\rm sh}}\bullet \rd \Omega/\int_{r_{\rm sh}}\rd \Omega$, and the domain of the integral is the surface slightly outside the shock. Here, $\langle A\rangle_{\rm sh}$ for the LS EOS is the mass number of a representative heavy nucleus, while that for the FS EOS is the average mass number of the nuclei in the NSE. For both EOSs, $\langle X_A\rangle_{\rm sh}$ is the mass fraction of the heavy nuclei.

The difference in the nuclear composition seems to play a key role in creating the difference in the entropy gradient. The heavy nuclei and alpha particles are finally dissolved into nucleons and consume the shock energy. The heavy nuclei and alpha particles in the LS model are more and less abundant than in the FS model, respectively. Hence, which model consumes more energy is not apparent. In order to estimate the loss of shock energy, figure \ref{fig:composition} also shows the total nuclear binding energy per nucleon of the accreting matter. Here, the binding energy of the heavy nuclei per nucleon is approximately set to $8.8\,{\rm MeV}$, the value for ${}^{56}\rm Fe$, because the binding energy of the nuclei around ${}^{56}\rm Fe$ is insensitive to the mass number. The binding energy of the alpha particle per nucleon is $7.1\,{\rm MeV}$. Therefore, the total binding energy per nucleon estimated here is $\langle X_A\rangle_{\rm sh}\times 8.8\,{\rm MeV} + \langle X_\alpha\rangle_{\rm sh}\times 7.1\,{\rm MeV}$. The figure indicates that the total binding energy is slightly higher for the LS model. It implies that more shock energy is lost due to photodissociation during that period, and the shock is weakened more rapidly in the LS model. This seems to produce the steeper negative radial gradient of the entropy to drive the prompt convection: the Brunt--V\"ais\"al\"a frequency, i.e., the convective growth rate, gets larger in the LS model. This appears to be the origin of the strong prompt convection in the LS model.

\begin{figure}[ht!]
\centering
\includegraphics[width=\hsize]{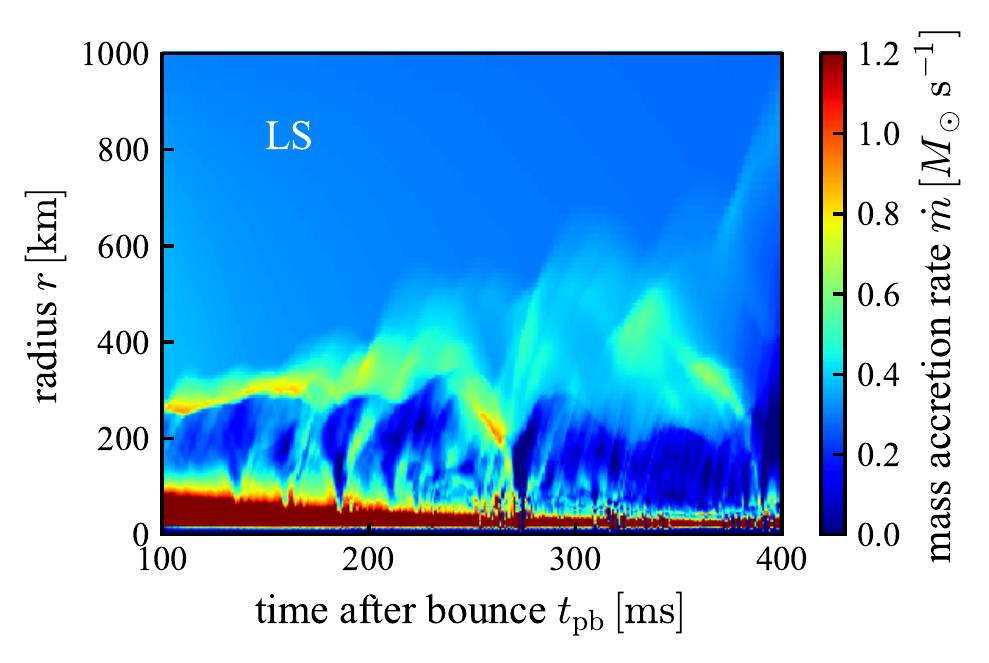}
\caption{The mass accretion rate $\dot{m}$ in the LS model as a function of the time and radius. This is evaluated from the angle-averaged density and radial velocity. \label{fig:LSaccretion}}
\end{figure}

Although the mass accretion rate at the shock $\dot{M}_{\rm sh}$ shown in figure \ref{fig:timescale} is smooth, that inside the shock is modulated by the shock motion. Precisely, the mass accretion rate at the shock is measured slightly outside the shock, where the flow is smooth. Figure \ref{fig:LSaccretion} shows the mass accretion rate $\dot{m}$ in the LS model as a function of time and radius. Here, the mass accretion rate is measured at a constant radius: $\dot{m} := - 4\pi r^2 \langle \rho \rangle \langle v^r \rangle$, where $\langle \bullet \rangle := \int_r \bullet \rd \Omega/\int_r \rd \Omega$ means the angular average. The shock significantly recedes at $t_{\rm pb}\sim 250\,{\rm ms}$. This results in significant mass accretion inside the shock at that time in figure \ref{fig:LSaccretion}. Owing to the released gravitational energy of this strong accretion, the $\nue$ luminosity increases momentarily as seen in figure \ref{fig:evolutions}.

\subsection{Structure of the PNS and the influence of the EOSs} \label{sec:PNSstructure}

\begin{figure}[ht!]
\plotone{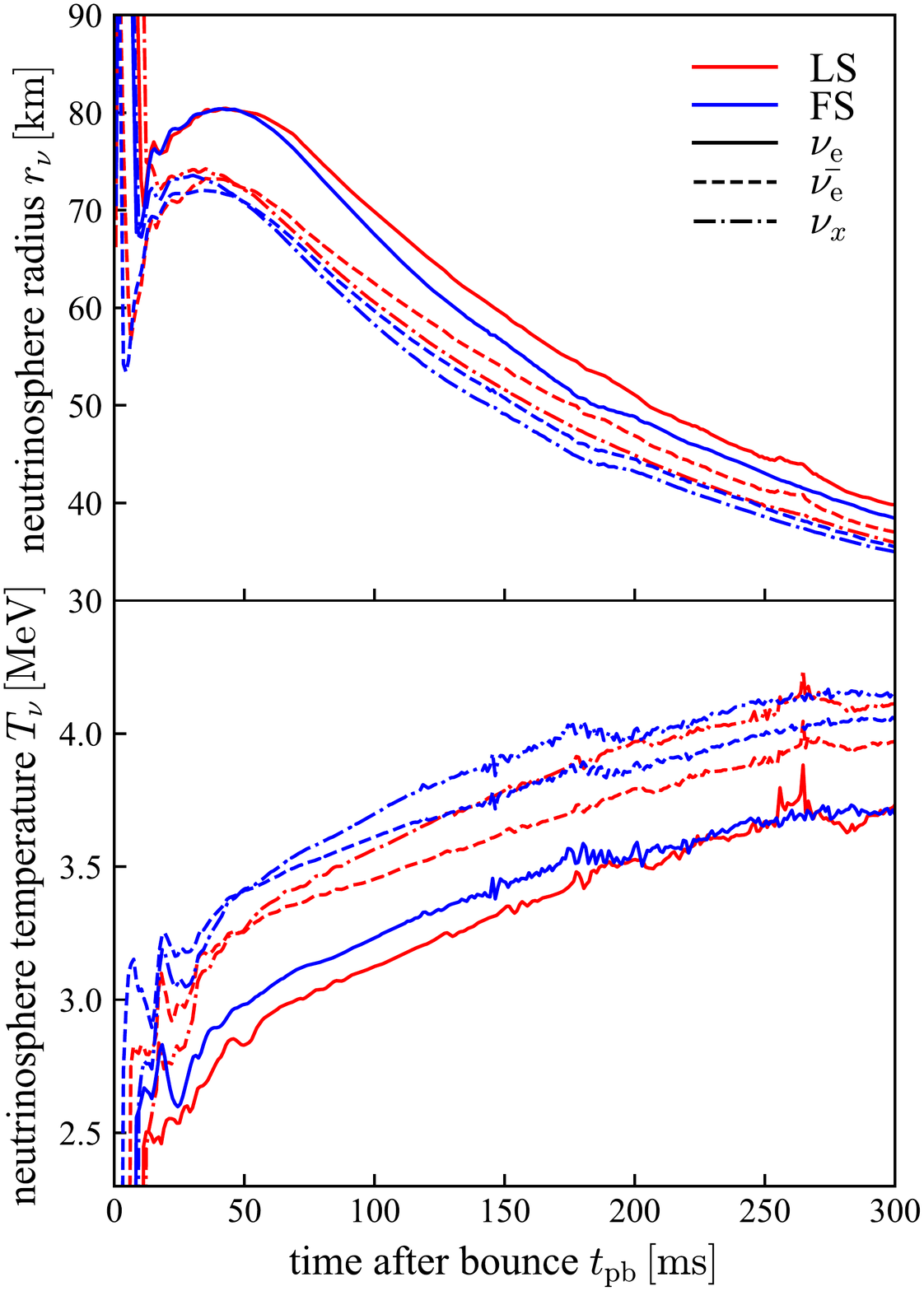}
\caption{The evolutions of the neutrinosphere radii and temperatures defined in the text. The red and blue lines are for the LS and FS models, respectively; the solid, dashed, and dash--dotted lines represent $\nue$, $\nueb$, and $\nux$, respectively. The time is truncated $300\,{\rm ms}$ after the bounce to focus on the comparison between the EOSs. \label{fig:neusph_rad_temp}}
\end{figure}

Figure \ref{fig:neusph_rad_temp} displays the evolutions of the neutrinosphere radii and temperatures. The neutrinosphere radius is defined as the radius where the angle-averaged total (absorption$+$scattering) optical depth for the mean-energy neutrino is $2/3$, and the neutrinosphere temperature is the angle-averaged temperature at the neutrinosphere radius. Here, we attempt to illustrate the similarity of the neutrino luminosities and differences in the mean and rms energies of neutrinos between the LS and FS models in figure \ref{fig:evolutions}. The neutrinosphere temperature is a useful indicator of the neutrino mean energies. The neutrinosphere radius is slightly smaller for the FS model, but the neutrinosphere temperature is higher for the FS model. Because they compensate for by each other, the neutrino luminosities for both models are similar. The difference in the temperature explains the difference in the mean and rms energies between the two models.

\begin{figure*}[ht!]
\plotone{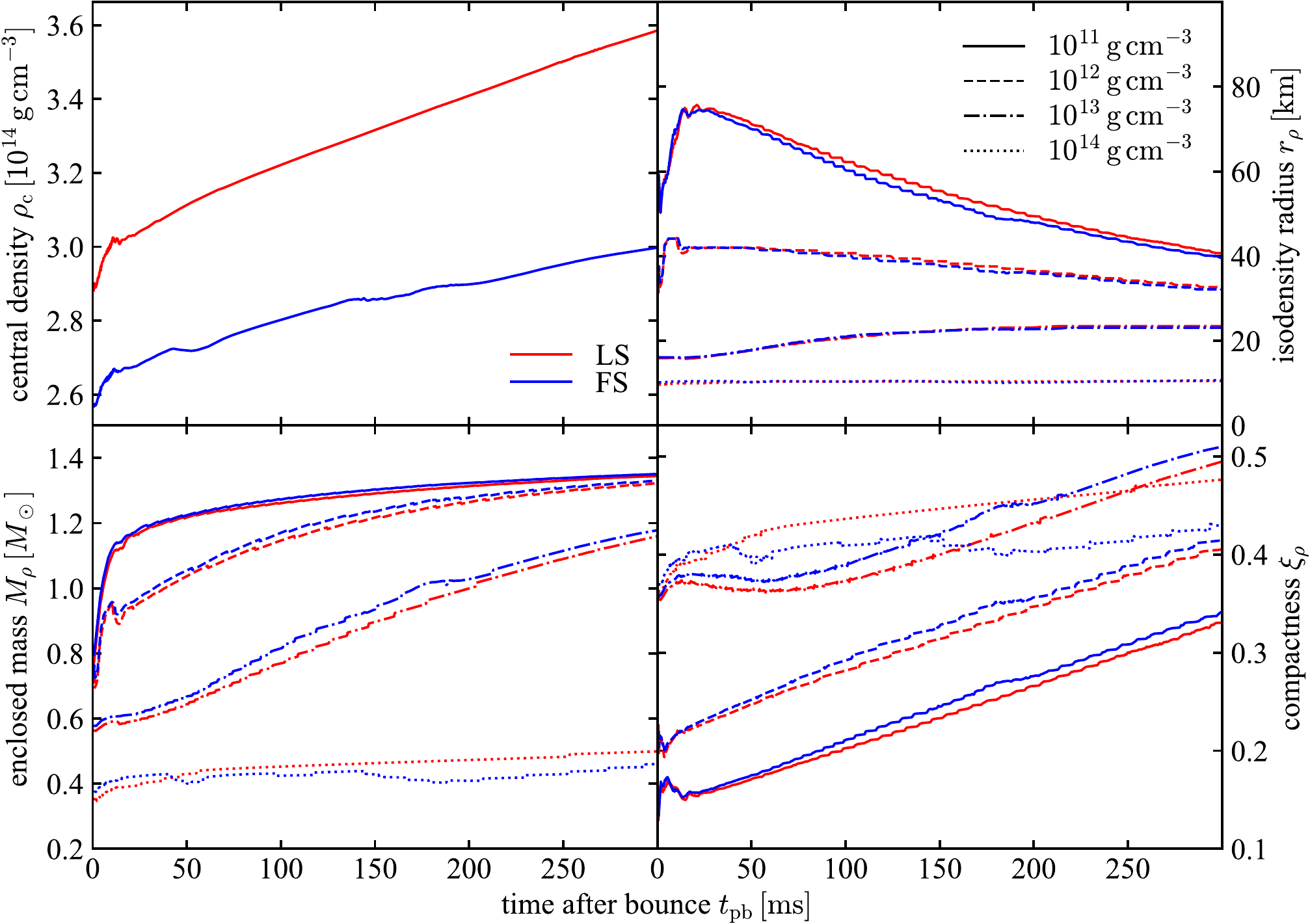}
\caption{The evolution of the central density and the PNS compactness for the LS (red) and FS (blue) models. The upper-left panel shows the central density evolution. The upper-right and lower-left panels indicate the radii and enclosed masses where the densities are $10^{11}$ (solid), $10^{12}$ (dashed), $10^{13}$ (dash--dotted), and $10^{14}$ (dotted) ${\rm g\,cm^{-3}}$. The lower-right panel indicates the compactness of each density defined by equation (\ref{eq:compactness}). Each quantity is calculated from the 1D averaged radial profiles. \label{fig:compactness}}
\end{figure*}

Although the LS EOS is known to be a soft EOS, the outer part of the PNS in the LS model is less compact than the FS model.
One may expect a smaller PNS radius, a smaller neutrinosphere radius, and hence a higher neutrinosphere temperature with the LS EOS compared with those with the FS EOS, which is a stiffer EOS. Figure \ref{fig:neusph_rad_temp}, however, shows the opposite result. To understand this, we show the time evolutions of the central density in figure \ref{fig:compactness}. The structure of the PNS, namely the radii, the enclosed masses, and the compactness at the angular-averaged densities of $10^{11}$, $10^{12}$, $10^{13}$, and $10^{14}\,{\rm g\,cm^{-3}}$, is also shown. Here, the compactness is defined as
\begin{equation}
\xi_\rho = \frac{M_\rho/M_\odot}{r_\rho/10\,{\rm km}}, \label{eq:compactness}
\end{equation}
where $M_\rho$ and $r_\rho$ are the enclosed mass and radius at the angle-averaged density $\rho$.
The central densities for the LS model are larger than those for the FS model, as expected. The radii and the enclosed masses at $\rho=10^{14}\,{\rm g\,cm^{-3}}$ are similar and larger for the LS model, respectively. Thus, the LS model shows higher compactness at this density. This is again consistent with the stiffness of the EOS. On the other hand, the radii and the enclosed masses at $\rho=10^{13}\,{\rm g\,cm^{-3}}$ are similar and smaller for the LS model, respectively. The resultant compactness for the LS model is lower than that for the FS model. At both $\rho=10^{12}$ and $10^{11}\,{\rm g\,cm^{-3}}$, the radii, the enclosed mass, and the compactness for the LS model are larger, smaller, and lower, respectively. The compactness for the LS model is higher only near the nuclear densities.

\begin{figure}[ht!]
\plotone{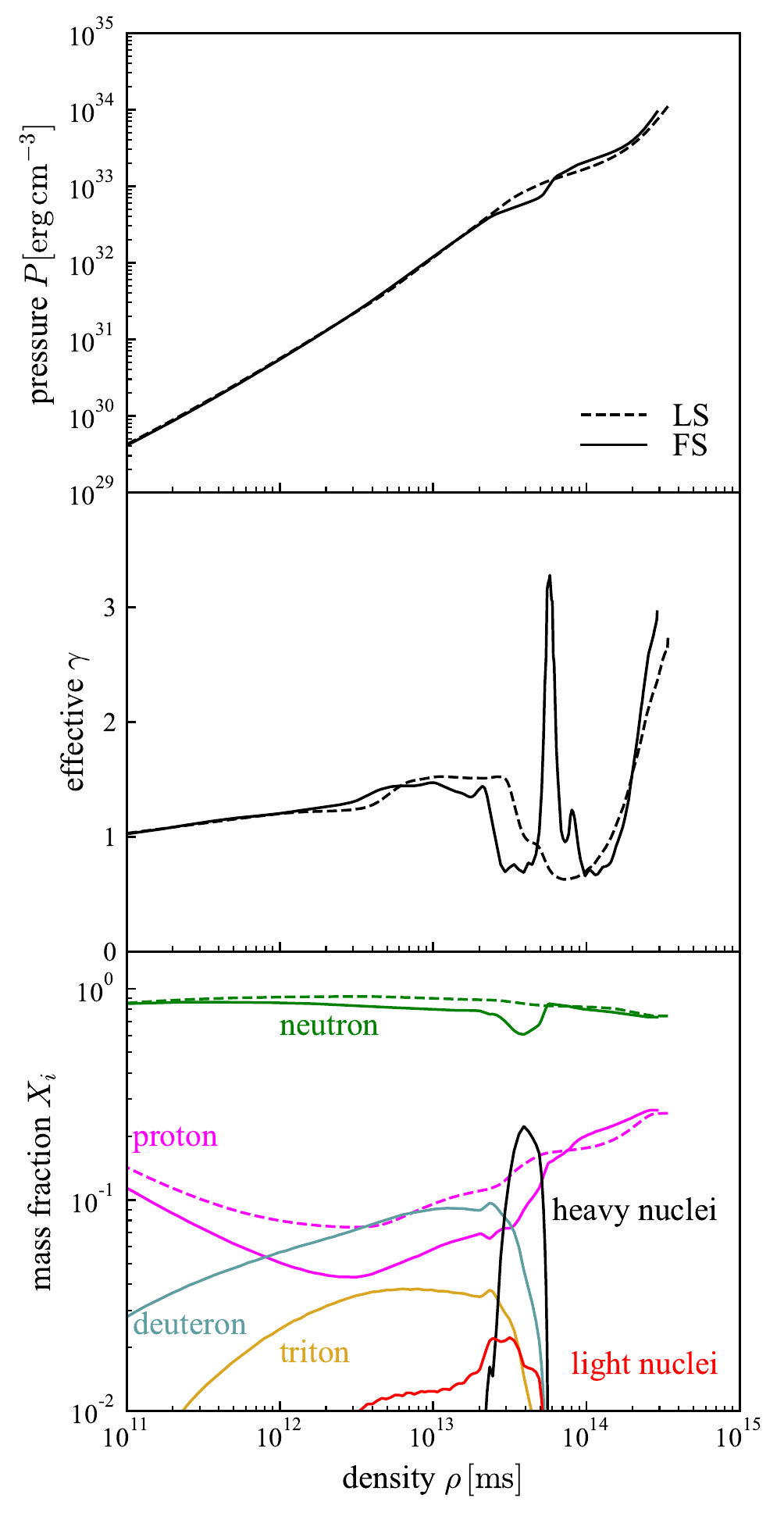}
\caption{The angular-averaged thermal properties and compositions vs. the density at $t_{\rm pb} \simeq 200\,{\rm ms}$ for the LS (dashed) and FS (solid) models. The top panel shows the relation between the density and the pressure, and the middle panel indicates the effective adiabatic index $\gamma$ (see the text). The bottom panel displays the corresponding composition: green for the neutron, magenta for the proton, cyan for the deuteron, yellow for the triton, red for the light nuclei, and black for the heavy nuclei. The light nuclei here are the nuclei with $Z<6$ other than deuteron, triton, helion, and alpha particle; the heavy nuclei are nuclei with $Z \ge 6$. For the FS EOS, the mass fractions of the helion and alpha particle are less than $10^{-2}$. \label{fig:rhopcomp}}
\end{figure}

The higher compactness in the FS model originates from the lower stiffness of the EOS at the densities around $10^{13}\,{\rm g\,cm^{-3}} \la \rho < 5\times 10^{13}\,{\rm g \,cm^{-3}}$. We show the pressure versus the density in the top panel of figure \ref{fig:rhopcomp}. The pressures for both EOSs are almost identical at densities lower than $\sim 10^{13}\,{\rm g\,cm^{-3}}$ and higher for the FS EOS at densities higher than $\sim 10^{14}\,{\rm g\,cm^{-3}}$. At around $\sim 4\times 10^{13}\,{\rm g\,cm^{-3}}$, the pressure is lower for the FS EOS. We show an effective adiabatic index $\gamma := \partial \ln P/\partial \ln \rho$ in the middle panel, although the ``stiffness'' of the EOS is determined by the adiabatic index $(\partial \ln P/\partial \ln \rho)_{s}$. Here, the entropy is not necessarily constant for the differentiation, and hence what is shown in the middle panel is nothing but the slope of the top panel. We use this effective $\gamma$ as an indicator of the stiffness. For densities higher than $\sim 5\times10^{13}\,{\rm g\,cm^{-3}}$, the effective $\gamma$ is higher for the FS EOS, indicating that the FS EOS is stiff. However, for density between $\sim 10^{13}\,{\rm g\,cm^{-3}}$ and $\sim 5\times10^{13}\,{\rm g\,cm^{-3}}$, the effective $\gamma$ is lower for the FS EOS. This softness leads to the more compact structure of the PNS at the region where the density is lower than $10^{13}\,{\rm g\,cm^{-3}}$ for the FS model as shown in figure \ref{fig:compactness}.

The mass accretion rate might influence the PNS compactness, but it is minor. If the mass accretion rate is low, the thermal energy provided to the PNS is low, and hence the PNS becomes compact \citep{2019ApJS..240...38N}. Although the mass accretion rate of the FS model is very similar to or slightly higher than that of the LS model until $\sim 200\,{\rm ms}$ after the bounce as shown in figure \ref{fig:timescale}, the PNS in the FS model is more compact. This implies that the effects of mass accretion rate is minor compared to that of the nuclear composition.

The low stiffness at the densities discussed above for the FS EOS originates from the difference in the composition. In the bottom panel of figure \ref{fig:rhopcomp}, we show the mass fractions of nuclei which are larger than $1\%$. In the high-density region $\rho > 10^{14}\,{\rm g\,cm^{-3}}$, only neutrons and protons are constituents of matter. In the subnuclear density region, light and heavy nuclei start to appear in the FS model, while no light and heavy nuclei appear for the LS model. This is simply because the NSE is considered in the FS EOS and light and heavy nuclei can appear, while only a single representative heavy nucleus and alpha particle are considered in the LS EOS. When searching for the free energy minimum to construct the EOS, there are more degrees of freedom for the FS EOS than those for the LS EOS due to the degrees of freedom in the composition. Therefore, the achieved free energy is smaller for the FS EOS, and hence the pressure and the effective $\gamma$ are smaller.

A caveat is that the updated version of the FS EOS \citep{2017NuPhA.957..188F} might be stiffer than the FS EOS employed in this paper, though it is probably still softer than the LS EOS. The temperature in the region where heavy nuclei appear in the FS model is $\sim 20\,{\rm MeV}$. The heavy nuclei are dissolved into free nucleons with temperatures above $18\,{\rm MeV}$ in the updated FS EOS \citep{2017NuPhA.957..188F}. This effect reduces the degree of freedom of the heavy nuclei, and hence, the updated FS EOS might be stiffer than the FS EOS in this paper. Although this melting of the heavy nuclei is not included in our current simulation, the degree of freedom of the composition of the light nuclei is probably still enough to make the updated FS EOS softer than the LS EOS at subnuclear densities. It is not only the stiffness but also the opacity that might be affected by the nuclear composition in the EOS. One might think that the difference in the neutrino mean and rms energies between the LS and FS models shown in figure \ref{fig:evolutions} is attributable to the different $Y_{\rm p}$ because the reaction rate of the electron capture on protons is reduced with small $Y_{\rm p}$. The opacity of the electron capture reaction in our simulations, however, does not fully take the composition shown in figure \ref{fig:rhopcomp} into account. Although the reduction of the proton mass fraction is due to the presence of the light nuclei, the weak interaction rate is evaluated as if the light nuclei were dissolved into free nucleons as described in section \ref{sec:method}. Hence, the proton mass fraction that was used to evaluate the electron capture rate in the FS model is not as low as $Y_{\rm p}$ in figure \ref{fig:rhopcomp} actually.

Even if we use $Y_{\rm p}$ in figure \ref{fig:rhopcomp} to evaluate the electron capture rate of free protons, the effects on the mean and rms energies are still limited. The neutrinosphere for the mean-energy neutrinos is located at $\sim 10^{11}\,{\rm g\,cm^{-3}}$. At this density, the nuclei are almost dissolved, and the $Y_{\rm p}$ in the FS model is close to that in the LS model in figure \ref{fig:rhopcomp}. The mean and rms energies are mainly determined by the neutrinosphere as discussed so far, and hence, the mean-energy neutrinos are hardly affected by the presence of the light nuclei in this case. Besides, the mean and rms energies are also affected by the low-density region. In the low-density region, the nuclei are completely dissolved, and hence $Y_{\rm p}$ is similar for the LS and FS models. Because the optical depth is determined by the integral of the opacity outside the given radius, the location of the neutrinosphere is almost determined by the opacity in the region that has nothing to do with the $Y_{\rm p}$ reduction in figure \ref{fig:rhopcomp}. We also note that the mean and rms energies are modulated during the propagation from the neutrinosphere to the radius where the neutrino quantities are measured. This is because the neutrinos with energies higher than the mean still interact with matter outside the mean-energy neutrinosphere, although the mean-energy neutrinos are decoupled from matter. This modulation is also determined by the opacity in the lowdensity region.

In general, the existence of light and heavy nuclei can itself influence the opacity \citep[e.g.,][]{2008PhRvC..77e5804S, 2016EPJ...109...6002F, 2019ApJS..240...38N}. The cross sections for the weak interactions of light and heavy nuclei are different from those of the nucleons \citep{2016EPJ...109...6002F}. If we consider the opacity of the weak interactions to be consistent with the composition shown in figure \ref{fig:rhopcomp}, the mean and rms energies of neutrinos might be modified as discussed in \citet{2019ApJS..240...38N}. This effect is, however, beyond the scope of this paper.

\section{Neutrino distributinos} \label{sec:neutrinos}
In this section, we discuss the physical quantities related to the neutrino angular distributions. The neutrino flux, the first angular moment, is discussed in section \ref{sec:flux}. The Eddington tensor, the second angular moment, is discussed in section \ref{sec:eddington}. Because we solve the Boltzmann equations for neutrinos, we can provide the information of the angular distributions.

\subsection{Neutrino flux} \label{sec:flux}

\begin{figure*}[ht!]
\includegraphics[width=\hsize]{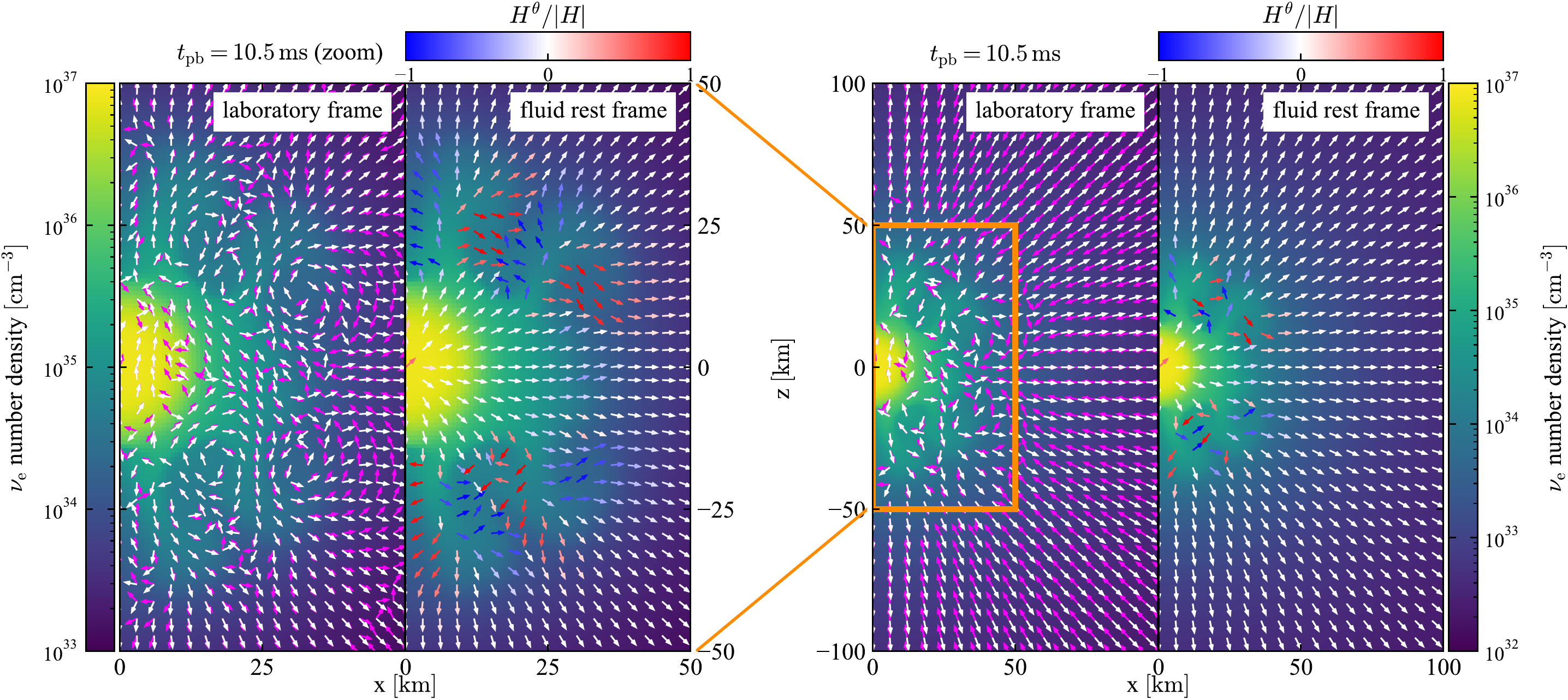}
\caption{The $\nue$ number flux and densities. The colormap shows the number density of $\nue$. For each panel, the white and magenta arrows in the left panel represent the direction of the neutrino flux in the laboratory frame and the matter velocity, respectively. On the other hand, the arrows in the right panel shows the direction of the neutrino flux in the fluid rest frame $H^i$ with their colors showing the lateral component: the reddish (bluish) colors correspond to the (counter)clockwise direction. The lengths of the arrows are normalized, and the absolute value of these vectors is not indicated here. The fluxes presented in this figure are measured at the mean energies.
Both panels are the snapshots at $t_{\rm pb}=10.5\,{\rm ms}$. The left panel is the zoom-in figure, and the right panel shows the larger region: the region shown in the left panel is indicated by the orange rectangle in the right panel. \label{fig:fluxarrow_nue_105}}
\end{figure*}

Figure \ref{fig:fluxarrow_nue_105} shows the number density and direction of the number flux of $\nue$ in both the laboratory and fluid rest frames at different scales at $t_{\rm pb} = 10.5\,{\rm ms}$. For the laboratory frame, the direction of fluid velocity is also shown. First, let us focus on the zoom-in snapshot (the left panel).

At the central $r\la 20\,{\rm km}$ region, the neutrino fluxes have various directions in the laboratory frame, while they are almost radially directed in the fluid rest frame. Because the central region is opaque, the neutrinos move in tandem with matter, and hence their directions in the laboratory frame are not uniform. On the other hand, the radial flux in the fluid rest frame comes from the diffusion of the neutrinos. The lateral motion seen in the laboratory frame is purely a consequence of neutrinos and matter comoving.
Note that due to the diffusion flux, the directions of the fluid velocity and the neutrino flux are slightly different. \cite{2019ApJ...878..160N} discussed that a precise evaluation of the flux in the employed code is difficult. The fixed-length arrows in figure \ref{fig:fluxarrow_nue_105} do not show the magnitude of the diffusion flux, and it is small at the very central region actually. Therefore, the evaluation of such small flux is perhaps not precise.

At the region where $20\,{\rm km} \la r \la 50\,{\rm km}$, neutrinos also move outward, but the lateral component alternately changes between clockwise (reddish arrows) and counterclockwise (bluish arrows) in the fluid rest frame. This is because of the convection and diffusion. The time considered here is the very early stage of the prompt convection. 
The region $20\,{\rm km} \la r \la 50\,{\rm km}$ is convectively unstable. The radial gradient of the $\nue$ number density is negative on average there. Due to the prompt convection, some fluid parcels rise and some sink in this region. The rising fluid parcels hence have a large $\nue$ number, whereas the sinking parcels have a small $\nue$ number. Thus, the finger-like pattern in the number density shown in the figure develops. Due to the difference in number density, diffusion flux emerges and neutrinos flow from rising parcels to sinking parcels in the fluid rest frame as clearly seen in the figure. In the laboratory frame, the neutrino flux is also affected by the matter velocity.

Although the flux in the inner region is determined by the diffusion and matter velocity, the neutrinos simply flow almost radially in the outer region. This is shown in the right panel of figure \ref{fig:fluxarrow_nue_105}. The zoom-in region shown in the left panel is indicated by the orange rectangle in the right panel. Because the outer region of the orange rectangle is relatively optically thin, the neutrinos freely stream, and the flux is radially directed.

\begin{figure*}[ht!]
\includegraphics[width=\hsize]{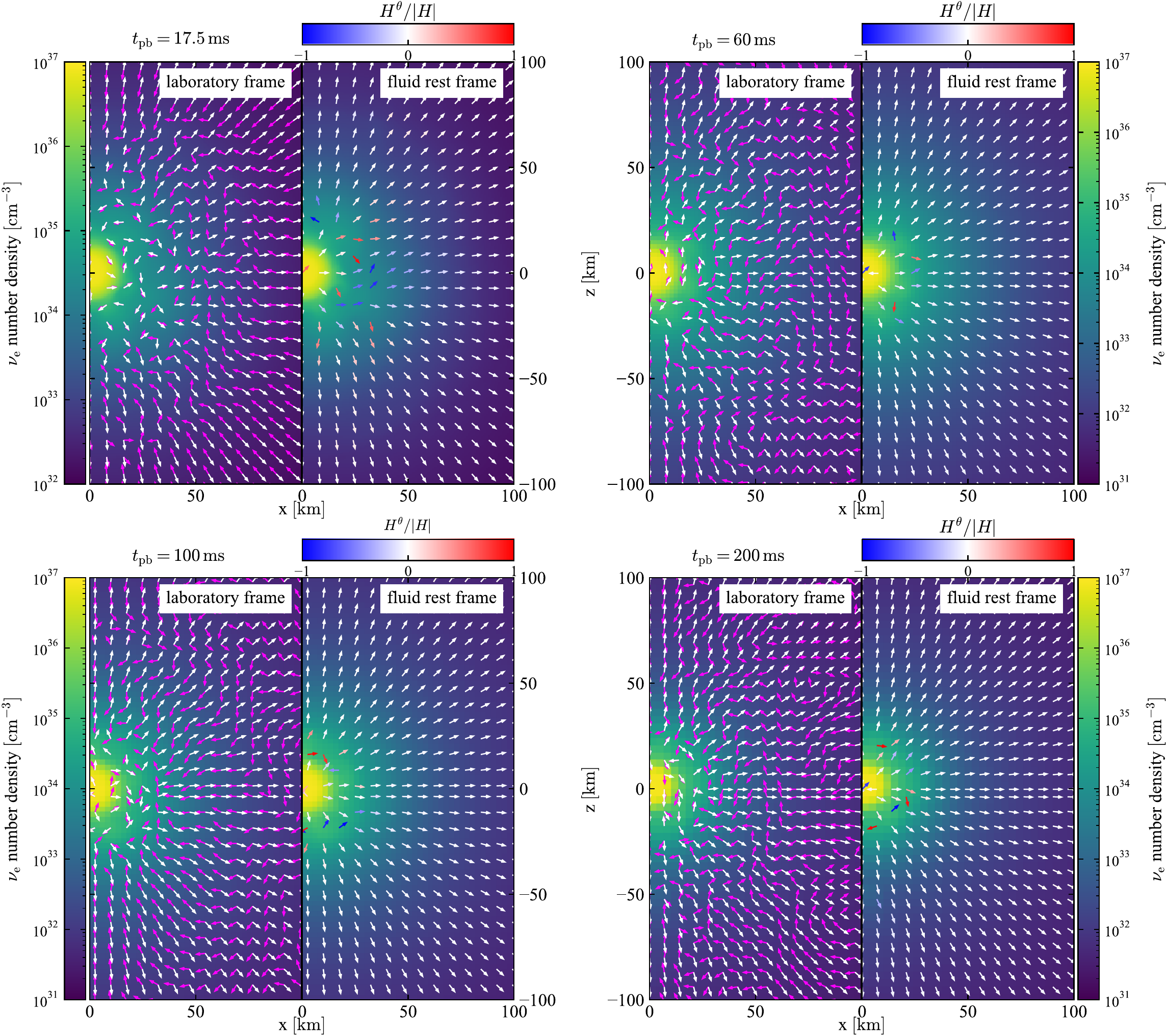}
\caption{The same as figure \ref{fig:fluxarrow_nue_105} except for the time of snapshots: $t_{\rm pb}=17.5\,{\rm ms}$ (upper left), $60\,{\rm ms}$ (upper right), $100\,{\rm ms}$ (lower left), and $200\,{\rm ms}$ (lower right). \label{fig:fluxarrow_nue_later}}
\end{figure*}

Figure \ref{fig:fluxarrow_nue_later} shows the $\nue$ number densities and fluxes at different times. Basically, the behavior of the neutrino flux is similar to the snapshot at $t_{\rm pb}=10.5\,{\rm ms}$: the direction is determined by the diffusion and matter velocity at the inner region and by the free-streaming at the outer region. However, at later times, the prompt convection diminishes and the region just outside the PNS is convectively stable. As a consequence, the neutrino number density is almost isotropic, and the diffusion flux is directed almost radially in the fluid rest frame. The flux in the laboratory frame is also almost radially directed but dragged by the matter velocity to some extent especially inside the PNS. The matter velocity inside the PNS originates from the PNS convection driven by the $\ye$ gradient.

\begin{figure*}[ht!]
\centering
\includegraphics[width=0.9\hsize]{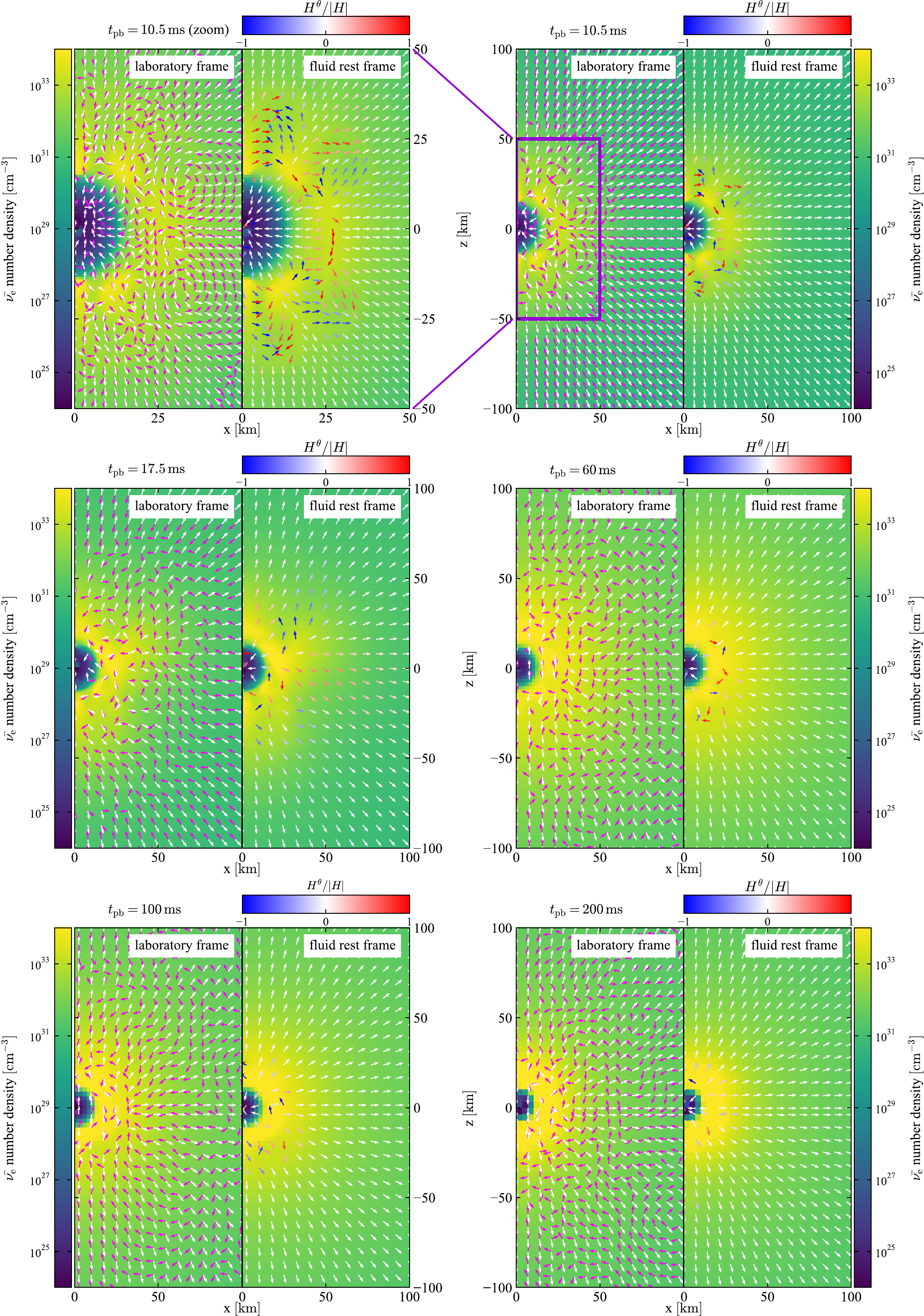}
\caption{The $\nueb$ number flux and densities. What is displayed is the same as figures \ref{fig:fluxarrow_nue_105} and \ref{fig:fluxarrow_nue_later} except that the neutrino species is not $\nue$ but $\nueb$. The snapshots at different scales and times are shown at once in this figure contrary to figures \ref{fig:fluxarrow_nue_105} and \ref{fig:fluxarrow_nue_later}. \label{fig:fluxarrow_nueb}}
\end{figure*}

The diffusion flux of $\nueb$ is different from that of $\nue$ due to the positive radial gradient of the $\nueb$ number density. Figure \ref{fig:fluxarrow_nueb} indicates the number density and direction of the number flux of $\nueb$. The flux in the upper-left panel shows a different pattern from the left panel of figure \ref{fig:fluxarrow_nue_105}. Although it is similar in that the lateral component changes alternately in the region where $20\,{\rm km} \la r \la 50\,{\rm km}$, the clockwise (reddish) and counterclockwise (bluish) pattern is opposite to the pattern seen in the $\nue$ flux. The flux in the fluid rest frame is directed inward around the center. These differences originate from the positive radial gradient of number density there. Due to the positive radial gradient, the sinking fluid parcels have a large $\nueb$ number, while the rising parcels have a small $\nueb$ number as shown in figure \ref{fig:fluxarrow_nueb}. The lateral component of the $\nueb$ diffusion flux hence shows the opposite sign to the $\nue$ diffusion flux.

The positive radial gradient of $\nueb$ comes from the degeneracy of $\nue$. The electron-type neutrinos are in $\beta$ equilibrium and degenerate at the center, and hence, their antiparticles are suppressed. Around the center, the matter density decreases, and the chemical potential of $\nue$ also decreases. The temperature increases around the center owing to the neutrino diffusion \citep{1981ApJ...251..325B, 1990RvMP...62..801B}. 
The decreasing chemical potential and increasing temperature make electron-type neutrinos nondegenerate, and hence electron-type antineutrinos start to appear. The degeneracy of $\nue$ decreases with the radius, then the number density of $\nueb$ increases with the radius.

Except for the different pattern in the $\nueb$ number density and the diffusion flux, the behavior of the $\nueb$ number flux is similar to that of the $\nue$ number flux. In the inner region, the flux is determined by the diffusion and the matter velocity. In the outer region, $\nueb$ streams freely. These behaviors are also seen in the snapshots at different times.

\begin{figure*}[ht!]
\centering
\includegraphics[width=0.9\hsize]{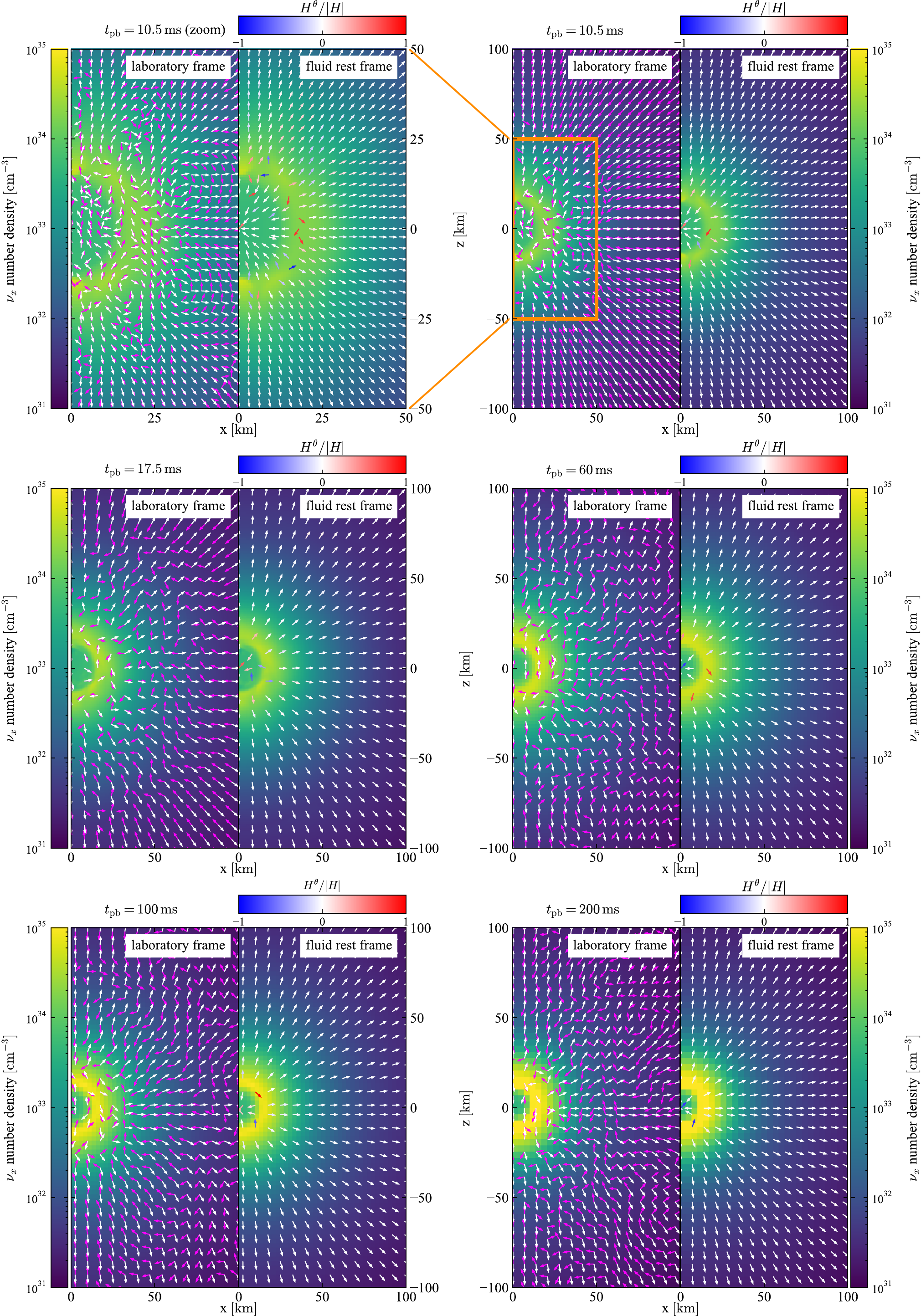}
\caption{The same as figure \ref{fig:fluxarrow_nueb}, except that the neutrino species is heavy-lepton type. \label{fig:fluxarrow_nux}}
\end{figure*}

Figure \ref{fig:fluxarrow_nux} displays the number density and direction of the number flux of $\nux$. Again, how the number flux of $\nux$ is determined is similar to that of $\nue$ and $\nueb$: the flux is determined by the diffusion and motion of matter in the optically thick region and the free-streaming in the optically thin region. The distribution of the flux of $\nux$ itself is similar to that of $\nueb$. Because the chemical potential is zero, the distribution of $\nux$ in the optically thick region is determined by the temperature. Therefore, the peak in the $\nux$ number density lies not in the center but around the center, and the radial gradient of the number density is positive around the center. Thus, the deformation by the convection results in the diffusion flux from the sinking fluid parcels to the rising parcels. However, compared with $\nueb$, the peak is located slightly closer to the center due to the zero chemical potential, and hence, the resultant pattern in the distribution and diffusion flux is also slightly different.

Finally, let us compare the neutrino fluxes of the LS and FS models briefly. Basically, the overall behavior is almost the same for both models, though we show no figures: the direction of the flux is determined by the diffusion and matter velocity at the inner region and by the free-streaming at the outer region. Due to the difference in EOSs, the distributions of the background fluid are different as discussed in section \ref{sec:dynamics}, and hence the resultant flux distributions are also different. Here we only focus on the early stage, when the turbulence is not fully developed. This is because the comparison at the late stage is not meaningful: the chaotic nature of the completely developed turbulence makes the flux patterns too different to be compared.

\begin{figure}[ht!]
\plotone{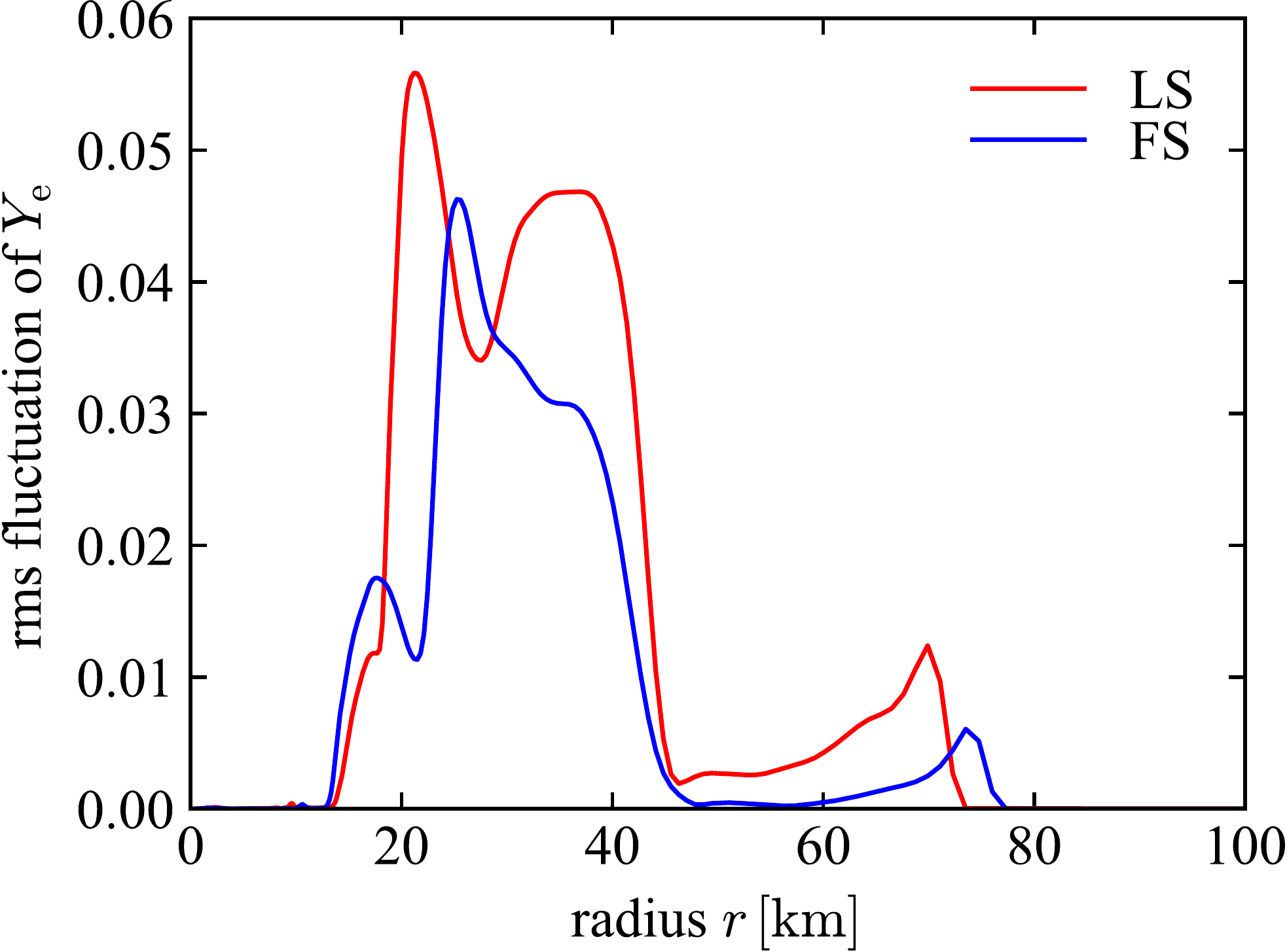}
\caption{The radial profiles of the rms fluctuation of the electron fraction $10.5\,{\rm ms}$ after the bounce. The red and blue lines represent the LS and FS models, respectively. \label{fig:RMSflucYe}}
\end{figure}

Figure \ref{fig:RMSflucYe} shows the radial profile of the angular rms fluctuation of $\ye$ in order to show how a large area is mixed by convection. Here, the angular rms fluctuation is defined as $\sqrt{\langle (\ye - \langle \ye \rangle)^2 \rangle}$ with $\langle \bullet \rangle := \int_r \bullet \rd \Omega / \int_r \rd \Omega$, where $r$ is given as a constant radius. The central region and the unshocked accretion flow are spherically symmetric, and hence the rms fluctuation is zero. On the other hand, the region $20\,{\rm km} \la r \la 50\,{\rm km}$ is mixed by the convection, and hence the rms fluctuation becomes large. The size of the region where the rms fluctuation is large in the FS model is smaller than that in the LS model because the prompt convection is stronger in the LS model.

\begin{figure}[ht!]
\plotone{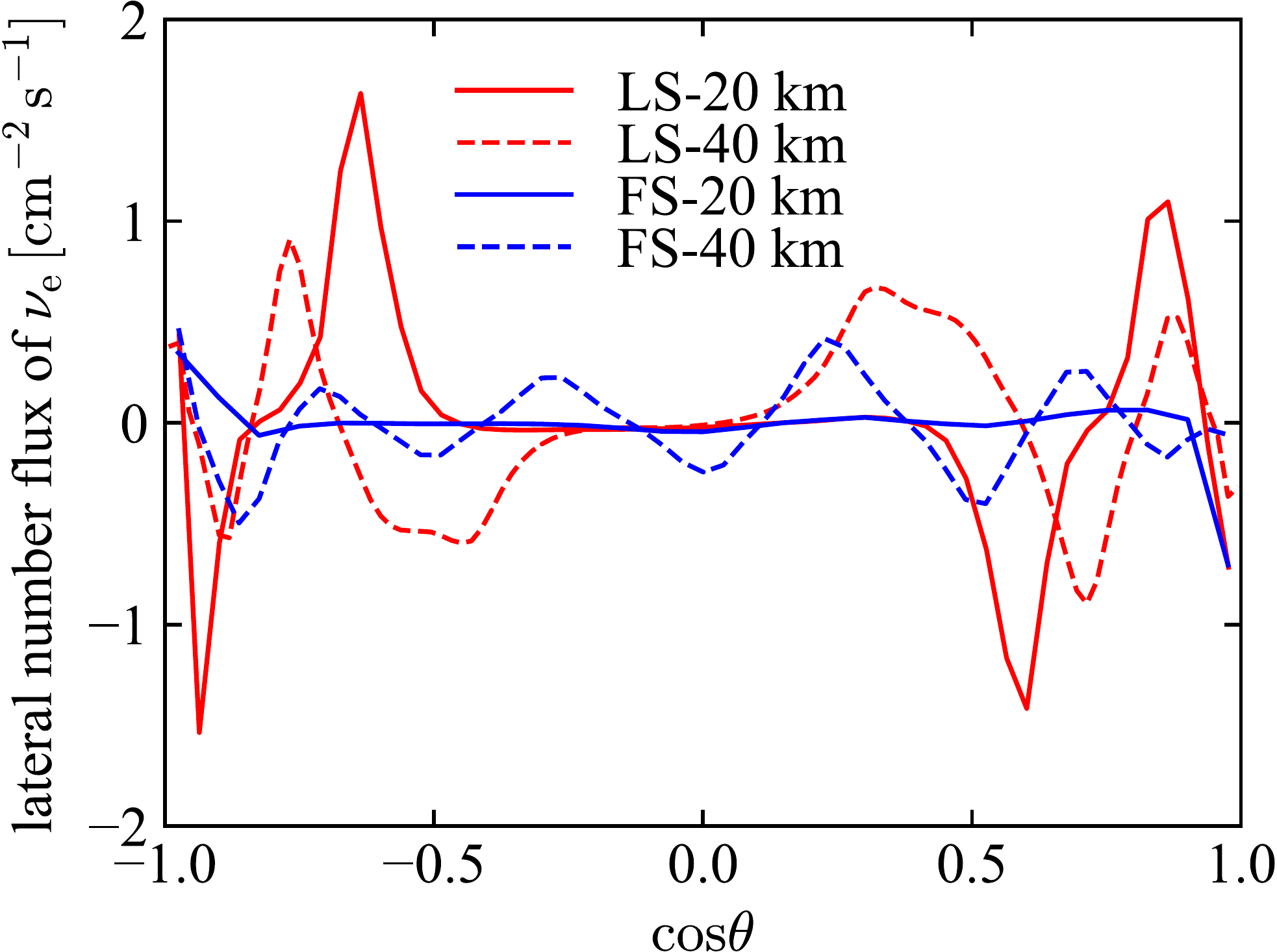}
\caption{The angular profiles of the $\theta$ component of the neutrino number flux in the fluid rest frame at different radii and models $10.5\,{\rm ms}$ after the bounce. The red and blue lines represent the LS and FS models, respectively. The solid and dashed lines are the angular profiles at $r=20\,{\rm km}$ and $r=40\,{\rm km}$, respectively. \label{fig:lateralflux}}
\end{figure}

As a result, the lateral flux by diffusion emerges from a larger region for the LS model as shown in figure \ref{fig:lateralflux}. We choose $r=20\,{\rm km}$ and $r=40\,{\rm km}$ for the radii where the angular profiles of the lateral flux are shown. This is because these radii are appropriate to see the difference in the strength of the prompt convection: the convective parcels in the FS model are almost confined between these radii, while those in the LS model overshoot these radii.
As expected, the diffusion flux in the FS model is much smaller than that in the LS model in those regions. In the region where the convection develops for both models, e.g., at the radius $r=30\,{\rm km}$, the orders of magnitude of the lateral diffusion fluxes are similar, though we do not show in the figure.

\subsection{Eddington tensor}
\label{sec:eddington}

\begin{figure*}[ht!]
\plotone{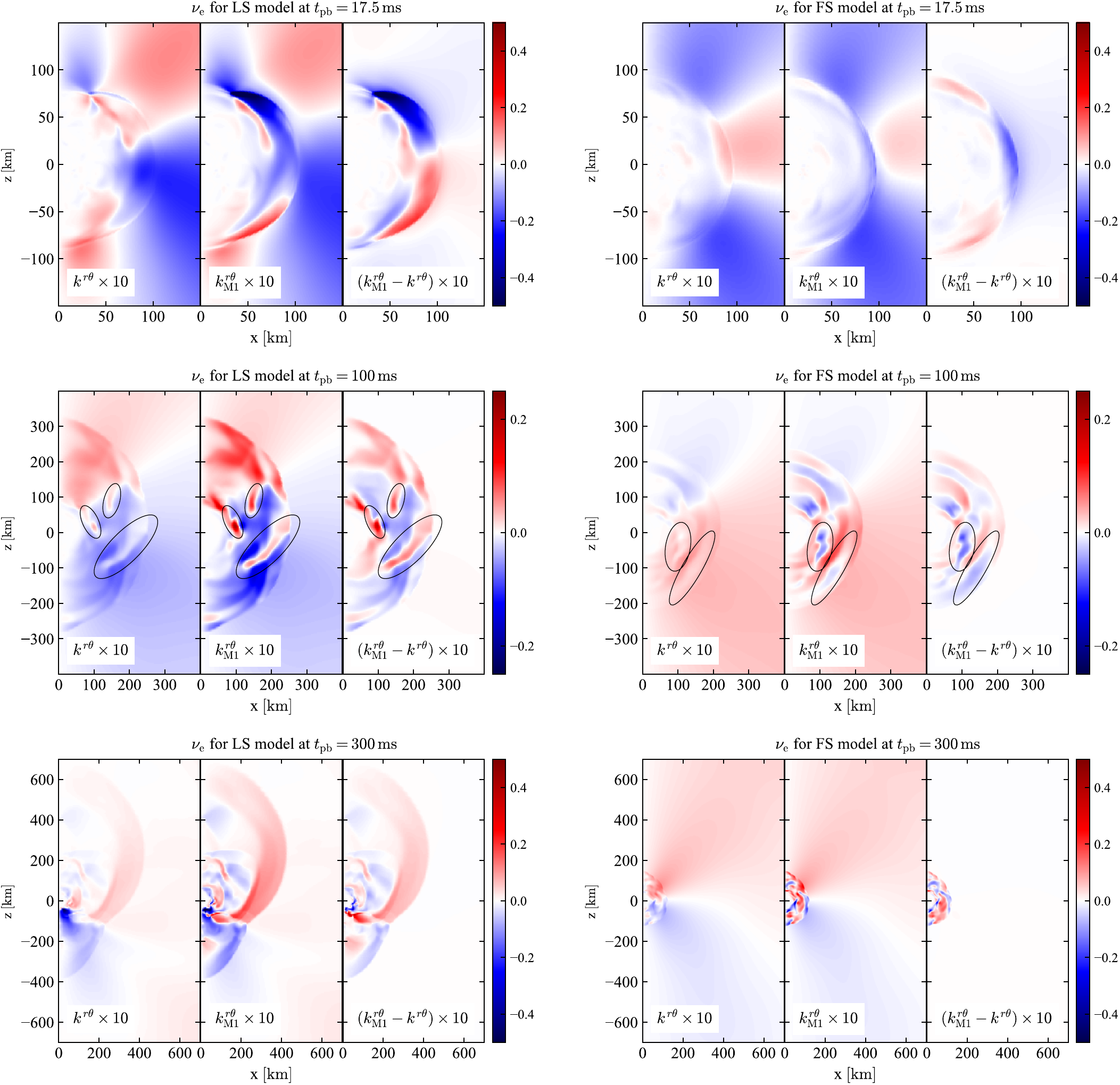}
\caption{The comparison of the $r\theta$ components of the Boltzmann- and M1-Eddington tensors for $\nue$. The left column is for the LS model, while the right column is for the FS model. The top, middle, and bottom rows are taken from the snapshots at $t_{\rm pb} = 17.5\,{\rm ms}$, $100\,{\rm ms}$, and $300\,{\rm ms}$, respectively. For each panel, the left and middle portions are the Boltzmann- and the M1-Eddington tensors, respectively; the right portion is the difference between them. They are all measured at the mean energies. The black circles in the middle panels indicate the filamental patterns discussed in the text. Note that the values are multiplied by a factor of 10 for the display. \label{fig:LSFSEddington}}
\end{figure*}

Let us now look into the Eddington tensor, the second moment of the distribution function. Here, we obey the definition of the Eddington tensor presented in \citet{2011PThPh.125.1255S},\footnote{The actual definition in \citet{2011PThPh.125.1255S} is slightly different from the definition presented here: the argument of the delta function is replaced from $\epsilon-\epsilon^\prime$ to $\epsilon^3/3-\epsilon^{\prime 3}/3$. We consider it is natural to choose the integral measure to the volume element of the momentum space because we do not treat the specific intensity of photons but the particle distribution function of neutrinos. This definition is the same as  \citet{2018ApJ...854..136N, 2019ApJ...872..181H}. The different definition does not affect the following discussions.} considering a $3+1$ decomposition of the spacetime. First, the second moment of the distribution function is defined as
\begin{eqnarray}
M^{\alpha\beta}(\epsilon) &:=& \int f \delta\left(\frac{\epsilon^3}{3} - \frac{\epsilon^{\prime 3}}{3}\right) p^{\prime \alpha} p^{\prime \beta} {\rm d}V_{p^\prime} \nonumber \\
&=& \frac{1}{\epsilon} \int f p^{\prime \alpha} p^{\prime \beta} \rd \Omega_p^\prime\Big|_{\epsilon^\prime=\epsilon},
\end{eqnarray}
where $\epsilon^\prime := - p^{\prime}\cdot e_{(0)}$ is the neutrino energy measured in the fluid rest frame \citep{1981MNRAS.194..439T}; ${\rm d}V_{p^\prime}$ is the invariant volume element of the momentum space. The integral in the second line is evaluated with the condition that the energy measured in the fluid rest frame is the constant $\epsilon$.
Next, we define the spatial--spatial and temporal--temporal projections of the second moment, which are nothing but the stress tensor and the energy density, respectively, as
\begin{equation}
P^{ij}(\epsilon) := \gamma^i{}_\alpha \gamma^j{}_\beta M^{\alpha\beta}(\epsilon)
\end{equation}
and
\begin{equation}
E(\epsilon) := n_\alpha n_\beta M^{\alpha\beta}(\epsilon),
\end{equation}
where $\gamma^i_\alpha$ and $n_\alpha$ are the spatial projection tensor and the normal vector to the spatial hypersurface, respectively. Finally, the Eddington tensor $k^{ij}(\epsilon)$ is defined by
\begin{equation}
k^{ij}(\epsilon) := \frac{P^{ij}(\epsilon)}{E(\epsilon)}. \label{eq:kBoltz}
\end{equation}
For the later convenience, we call this $k^{ij}(\epsilon)$ the ``Boltzmann-Eddington tensor".

We compare the Boltzmann-Eddington tensor and the Eddington tensor calculated from the M1-closure approximation. The M1-closure scheme gives the Eddington tensor from the energy flux and energy density of neutrinos to close the moment equations of the radiative transfer up to the first order. Hence, the Eddington tensor in the M1-closure scheme plays a key role in evolving the energy flux of neutrinos. Because a direct comparison of the results of the radiation-hydrodynamics simulation with the Boltzmann neutrino transport and those with the M1-closure scheme is not in the scope of this paper, we compare the Eddington tensors. Hereafter, we call the Eddington tensor calculated with the M1-closure prescription the ``M1-Eddington tensor" $k^{ij}_{\rm M1}(\epsilon)$.

Here, we follow the M1-closure scheme suggested in \citet{2011PThPh.125.1255S}. In the M1-closure scheme, two limiting cases are interpolated to obtain an approximate value of the stress tensor $P_{\rm M1}^{ij}(\epsilon)$:
\begin{equation}
P^{ij}_{\rm M1} ( \epsilon) := \frac{3\zeta(\epsilon) - 1}{2} P^{ij}_{\rm thin} (\epsilon) + \frac{3(1-\zeta(\epsilon))}{2} P^{ij}_{\rm thick} (\epsilon), \label{eq:pm1}
\end{equation}
where $P_{\rm thin}^{ij}$ and $P_{\rm thick}^{ij}$ are the optically thin and thick limits of the stress tensor, respectively; $\zeta(\epsilon)$ is the Eddington factor, which is defined as the eigenvalue of the Eddington tensor along the flux direction. The M1-closure scheme gives the Eddington factor by an analytic function of the flux factor $\bar{F}(\epsilon) := \sqrt{h_{\alpha\beta} H^\alpha(\epsilon) H^\beta(\epsilon)/J(\epsilon)^2}$:
\begin{equation}
\zeta(\epsilon) = \frac{3+4\bar{F}(\epsilon)^2}{5+2\sqrt{4-3\bar{F}(\epsilon)}}.
\end{equation}
Here, $J(\epsilon) : = u_\alpha u_\beta M^{\alpha\beta}(\epsilon)$ and $H^i(\epsilon) := - h^i_\alpha u_\beta M^{\alpha\beta}(\epsilon)$ are the energy density and flux in the fluid rest frame, respectively; $u^\alpha$ is the four-velocity of the fluid; $h_{\alpha\beta} := g_{\alpha\beta} + u_\alpha u_\beta$ is the projection tensor onto the fluid rest frame.
With this stress tensor $P_{\rm M1}^{ij}(\epsilon)$, we define the M1-Eddington tensor as $k_{\rm M1}^{ij}(\epsilon):=P_{\rm M1}^{ij}(\epsilon)/E(\epsilon)$

The optically thin and thick limits of the stress tensor are given as follows:
\begin{equation}
P^{ij}_{\rm thin}(\epsilon) := E(\epsilon) \frac{F^i(\epsilon)F^j(\epsilon)}{F(\epsilon)^2}
\end{equation}
at the optically thin limit and
\begin{equation}
P^{ij}_{\rm thick} (\epsilon) := J(\epsilon) \frac{\gamma^{ij}+ 4 V^i V^j}{3} + H^i(\epsilon) V^j + V^i H^j(\epsilon) \label{eq:pthick}
\end{equation}
at the optically thick limit. Here, some projected quantities are also utilized: $F^i(\epsilon) := - \gamma^i_\alpha n_\beta M^{\alpha\beta}(\epsilon)$ is the spatial-temporal projection, or the energy flux, in the laboratory frame; $V^i := u^i/u^0$ is the three-velocity of the fluid.

In the following, we evaluate the M1-Eddington tensor from the second moment of the distribution function $M^{\alpha\beta}(\epsilon)$: we first obtain the energy densities and fluxes $E(\epsilon)$, $J(\epsilon)$, $F^i(\epsilon)$, and $H^i(\epsilon)$ by projecting $M^{\alpha\beta}$ then calculate the stress tensor $P_{\rm M1}^{ij}(\epsilon)$ from equations (\ref{eq:pm1}--\ref{eq:pthick}). In the standard M1-closure scheme, however, we would not have the variables in both the laboratory and fluid rest frames. Usually, we only have the energy density and flux in the laboratory frame ($E(\epsilon)$ and $F^i(\epsilon)$) instead of the entire second moment. Then, we guess the energy density and flux in the fluid rest frame ($J_{\rm guess}(\epsilon)$ and $H_{\rm guess}^i(\epsilon)$), construct the stress tensor $P_{\rm M1,guess}^{ij}(\epsilon)$ via equation (\ref{eq:pm1}), construct the second moment $M_{\rm guess}^{\alpha\beta}(\epsilon)$ \citep[see equation (3.29) in][]{2011PThPh.125.1255S} from $E(\epsilon)$, $F^i(\epsilon)$, and $P_{\rm M1,guess}^{ij}(\epsilon)$, and obtain new guesses of $J_{\rm guess}(\epsilon)=u_\alpha u_\beta M_{\rm guess}^{\alpha\beta}(\epsilon)$ and $H_{\rm guess}^i(\epsilon)=-h^i_\alpha u_\beta M_{\rm guess}^{\alpha\beta}(\epsilon)$ in the fluid rest frame.
By iterating this correction numerically until the guess converges, we finally obtain the consistent energy density $J(\epsilon)$, flux $H^i(\epsilon)$, and stress tensor $P_{\rm M1}^{ij}(\epsilon)$. Although we compare this iteratively obtained M1-Eddington tensor and our M1-Eddington tensor obtained from the second moment, the difference is only $\la 10^{-4}$ at $t_{\rm pb} = 100\,{\rm ms}$. Therefore, we do not use the iteratively obtained stress tensor for the M1-Eddington tensor.

It is worth noting that equation (\ref{eq:pthick}) is derived from the perturbative expansion of the angular moments of the distribution function with respect to the local MFP. The second- or higher-order terms of the local MFP are ignored in the M1 prescription. Instead of including the higher-order terms in the semitransparent cases, the M1 scheme takes their effect into account by interpolating with the optically thin limit. Besides, some terms are ignored in equation (\ref{eq:pthick}) compared to the original expression \citep[equation (6.19) in][]{2011PThPh.125.1255S}. This is just because such prescription makes the Boltzmann- and M1-Eddington tensors match better.

The M1-Eddington tensor shows overestimated filamental patterns compared with the Boltzmann-Eddington tensor. The Boltzmann- and M1-Eddington tensors and their difference for $\nue$ are displayed in figure \ref{fig:LSFSEddington}. In the circles in the middle-left panel of figure \ref{fig:LSFSEddington} ($\nue$ for the LS model at $t_{\rm pb}=100\,{\rm ms}$), red or white filaments are seen on the bluish background. Although the filaments are almost white or pale red ($k^{r\theta} \ga 0$) in the panel for the Boltzmann-Eddington tensor, the filaments in the panel for the M1-Eddington tensor are red ($k_{\rm M1}^{r\theta} > 0$). The circles in the middle-right panel ($\nue$ for the FS model at $t_{\rm pb} = 100\,{\rm ms}$) show white or blue filaments on the reddish background. These filaments in the Boltzmann-Eddington tensor are white ($k^{r\theta} \sim 0$), while those in the M1-Eddington tensor are blue ($k_{\rm M1}^{r\theta} < 0$). Similar overestimated filamental patterns are also found in other regions or other panels.

This overestimation comes from the limitation of the approximation. The central idea of the M1-closure scheme is the interpolation between $P_{\rm thick}^{ij}$ and $P_{\rm thin}^{ij}$ to approximate $P^{ij}$: the inequality $|P_{\rm thick}^{r\theta}| \le |P^{r\theta}| \le |P_{\rm thin}^{r\theta}|$ is assumed. In reality, $|P_{\rm thin}^{r\theta}| \le |P_{\rm thick}^{r\theta}|$ holds in the overestimated filaments, and hence, the assumption of the M1-closure relation is violated.

The overestimation of $|P_{\rm thick}^{r\theta}|$ originates from the ignored higher-order terms of the local MFP. 
The origin of the filamental shapes seen in figure \ref{fig:LSFSEddington} is the lateral matter velocity: the distribution of the neutrinos are slightly distorted by the lateral matter velocity in the semitransparent region via the scattering or the emission. 
In the M1-closure scheme, the matter motion is encoded through $V^i$ in equation (\ref{eq:pthick}). Equation (\ref{eq:pthick}) includes up to first-order terms of the local MFP, and higher-order terms are ignored.
In the filamental patterns in figure \ref{fig:LSFSEddington}, the $H^r V^\theta$ term becomes too large because the matter moves laterally in the semitransparent regions: $H^r$ is no longer the first order of the MFP and $V^\theta$ is nonnegligible. This term should be canceled by the higher-order terms, but they are ignored in this formulation. Hence, $|P_{\rm thick}^{r\theta}|$ becomes too large for the assumption of the M1-closure scheme to hold. Similar discussions are also presented in \citet{2019ApJ...872..181H}.

\citet{2017ApJ...847..133R} showed that the limited angular resolution of the $S_N$ solver makes the off-diagonal component of the Eddington tensor underestimated. Thus, higher-resolution simulations probably show less difference between the Boltzmann- and M1-Eddington tensors. The possible violation of the assumption of the M1-closure relation, the inequality $|P_{\rm thick}^{r\theta}| < |P_{\rm thin}^{r\theta}|$, is still problematic, however.

One of the peculiar features in figure \ref{fig:LSFSEddington} is the very (negative) large value of $k_{\rm M1}^{r\theta}$ colored in dark blue in the top-left panel (LS model at $t_{\rm pb} = 17.5\,{\rm ms}$). This panel corresponds to the first phase of the prompt convection, and the shock expanding from the equator rides over the shock in the upper region.
A significant lateral motion to the negative $\theta$ direction arises there, and hence, such matter motion produces a large lateral flux to the negative $\theta$ direction. This lateral flux results in the large $F^r F^{\theta}$ and hence the large $P_{\rm thin}^{r\theta}$. Besides, the large $V^\theta$ makes $P_{\rm thick}^{r\theta}$ large. Both of them result in the large $k^{r\theta}_{\rm M1}$. Note that the large $F^\theta$ does not necessarily result in the large $P^{r\theta}$. This is because the $P^{r\theta}$ is related to the quadrupole moment of the distribution function, while the $F^i$ is the dipole moment of the distribution.

\begin{figure*}[ht!]
\plotone{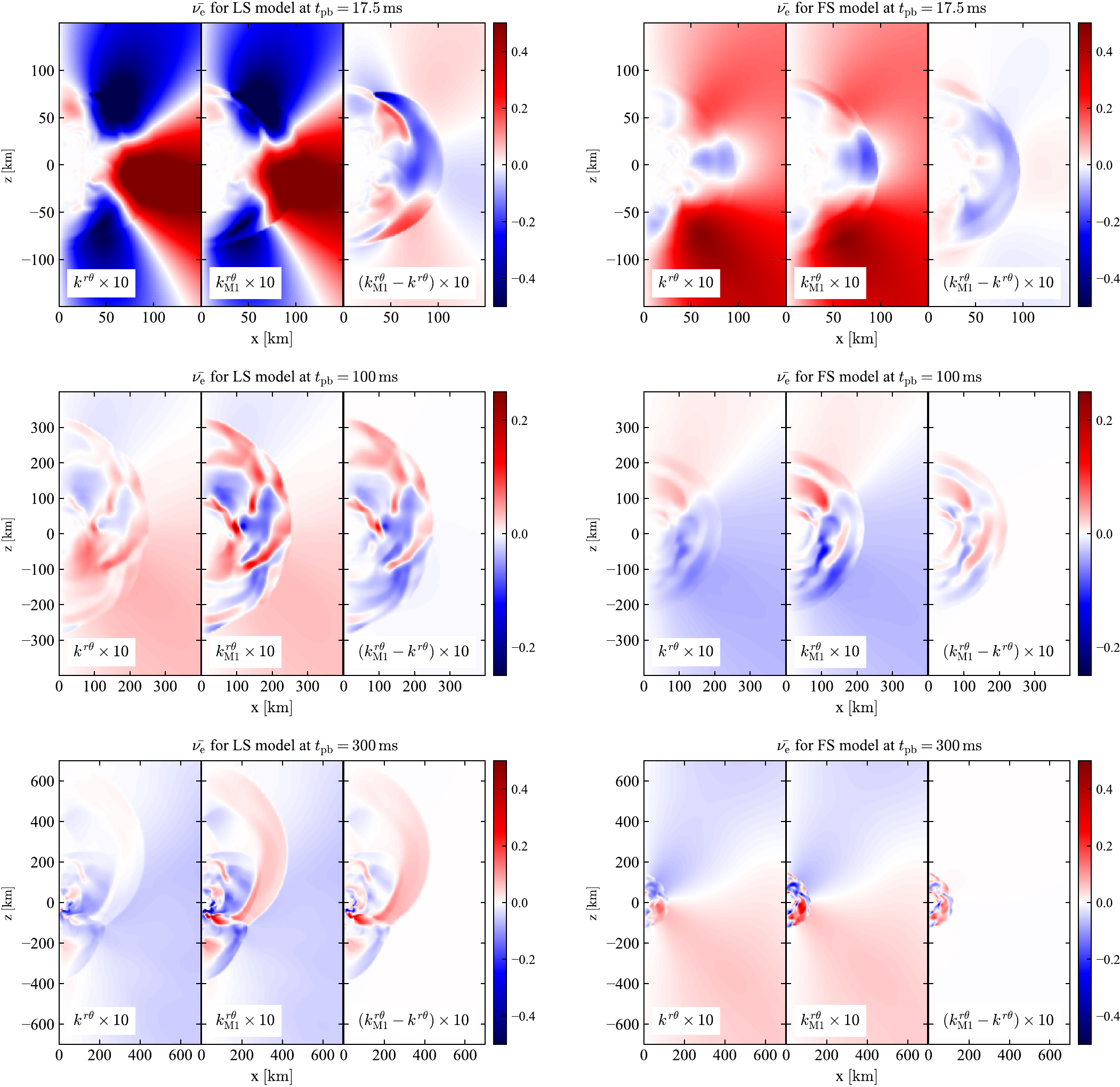}
\caption{The same as figure \ref{fig:LSFSEddington} except that the displayed neutrino species is $\nueb$. \label{fig:LSFSEddingtonnueb}}
\end{figure*}

\begin{figure*}[ht!]
\plotone{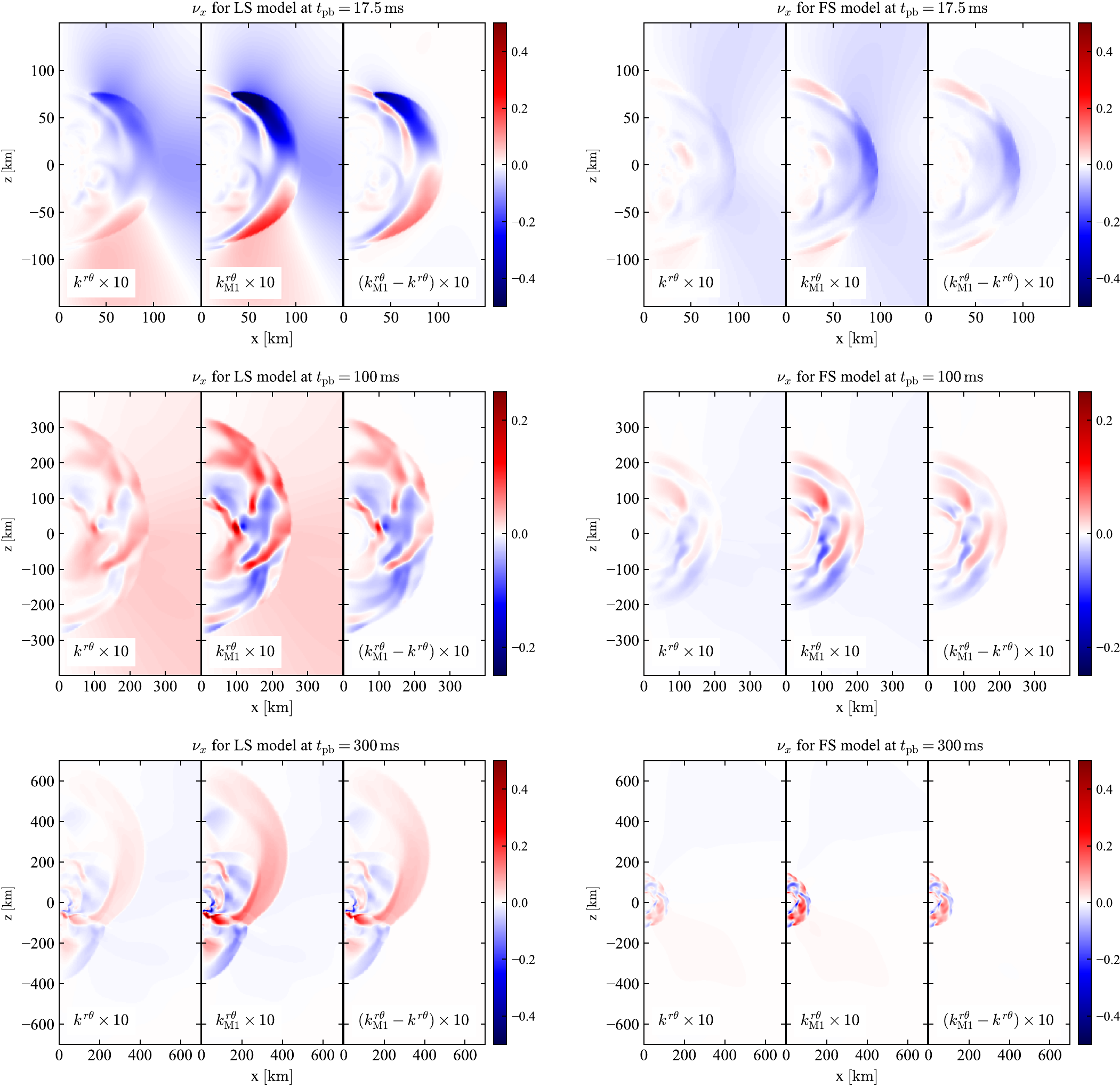}
\caption{The same as figures \ref{fig:LSFSEddington} and \ref{fig:LSFSEddingtonnueb} except that the displayed neutrino species is $\nux$. \label{fig:LSFSEddingtonnux}}
\end{figure*}

The Eddington tensors of $\nueb$ displayed in figure \ref{fig:LSFSEddingtonnueb} show similar background and opposite filamental patterns to those of $\nue$. The basic pattern of the signature, which is the background pattern behind the filamental patterns, is opposite between $\nue$ (figure \ref{fig:LSFSEddington}) and $\nueb$ (figure \ref{fig:LSFSEddingtonnueb}). For example, the signs in the equatorial and northern/southern regions of the upper-left panel of figure \ref{fig:LSFSEddingtonnueb} ($\nueb$ for LS model at $t_{\rm pb}=17.5\,{\rm ms}$) are positive and negative, respectively, contrary to those of figure \ref{fig:LSFSEddington};
the sign in the upper region of the middle-left panel of figure \ref{fig:LSFSEddingtonnueb} ($\nueb$ for LS model at $t_{\rm pb}=100\,{\rm ms}$) is negative while that of figure \ref{fig:LSFSEddington} for $\nue$ is positive. On the other hand, the filamental patterns for $\nue$ and $\nueb$ are similar. The filaments that are enclosed by the black circles in the middle panels of figure \ref{fig:LSFSEddington} ($\nue$ at $t_{\rm pb}=100\,{\rm ms}$) are also found in the middle panels of figure \ref{fig:LSFSEddingtonnueb} ($\nueb$ at $t_{\rm pb}=100\,{\rm ms}$).

The $r\theta$ component of the Eddington tensor is determined by the emission from the optically thick region, whose signature is different between $\nue$ and $\nueb$, and the motion of matter, whose signature is the same between the two neutrino species.
The different signs of the background pattern discussed above originate from the degeneracy of $\nue$ discussed in section \ref{sec:flux}. In the optically thick region, the number density of $\nueb$ is large in the region where the $\nue$ number density is small, and the lateral component of the diffusion flux of $\nueb$ is the opposite to that of $\nue$. This is imprinted not only in the flux but also the $r\theta$ component of the Eddington tensor. For the semitransparent to optically thin region, though $\nue$ is no longer degenerate, the pattern generated in the optically thick region is transported and shows the same signature as the inner regions. The background pattern is hence determined by the emission from the optically thick region. On the other hand, the filamental patterns are induced by the motion of matter. Since the motion of matter is common for both neutrino species, the filamental patterns for $\nue$ and $\nueb$ are similar.

The Eddington tensors for $\nux$ are shown in figure \ref{fig:LSFSEddingtonnux}. The filamental pattern is similar to $\nue$ and $\nueb$. This is because the motion of matter again determines the signature there. However, the pattern in the other regions does not have a clear (anti)correlation with $\nue$ and $\nueb$. Because the number density is solely determined by the temperature and has nothing to do with the chemical potential of $\nue$, the resultant pattern is again not so related to that of $\nue$ and $\nueb$.

\section{Summary and Conclusions} \label{sec:concl}

In this paper, we have discussed the postbounce dynamics and the neutrino properties of the models with different EOSs. They are summarized as follows:
\begin{enumerate}
\item The model with the LS EOS shows shock revival, while that with the FS EOS does not. The neutrino luminosities, energies, heating rates, the mass accretion rates, and total energies in the gain region of the two models are similar, while the gain mass and strength of the turbulence are different. This originated from the difference in the nuclear composition of the accreting matter. The estimated binding energy is larger for the LS model, and hence more energy is consumed to photodissociate them. It results in a steeper entropy gradient, which drives stronger prompt convection. It enhances the later neutrino-driven convection.

\item The structure of the PNS implies the importance of the composition. The central region of the PNS in the LS model is more compact than that in the FS model. However, the outer region in the LS model is slightly less compact than that in the FS model. This is understood from the effective stiffness. The nuclear composition under the NSE has more degrees of freedom than that under the SNA. Because the FS EOS considers the NSE, the free energy in the FS model is lower than that in the LS model. Therefore, the pressure gets lower in the FS model, and hence, the FS EOS at subnuclear densities is softer than the LS EOS.

\item The neutrino flux is determined by the diffusion of the neutrinos, whose direction is along the gradient of the neutrino number density, and the matter velocity via the Lorentz transformation. The $\nue$ number density decreases with the radius, and hence the diffusion flux is directed from the rising fluid parcels to the sinking parcels when the prompt convection develops. On the other hand, the $\nueb$ number density increases with the radius in the region where prompt convection occurs. This is because the $\nue$ number density and hence the degeneracy of $\nue$ decreases with the radius. Therefore, the diffusion flux is directed from sinking to rising fluid parcels. For the heavy-lepton-type neutrinos, its number density is determined by the zero-chemical-potential thermal distribution: it increases with the temperature. Therefore, the diffusion flux is similar to that of the electron-type antineutrinos, although is slightly different because the peak in the number density lies in the convectively unstable region.

\item We compared the Boltzmann- and M1-Eddington tensors and found that the effect of the matter velocity on the Eddington tensor is overestimated in the M1-Eddington tensor. This is mainly because the higher-order terms of the local MFP ignored in equation (\ref{eq:pthick}) is too large for the M1 approximation to hold. The Eddington tensor itself is affected by both the diffusion of neutrinos and the matter velocity. The contribution of the neutrino diffusion is opposite between $\nue$ and $\nueb$ due to the $\nue$ degeneracy. The contribution of the diffusion of $\nux$ has nothing to do with $\nue$ and $\nueb$ because the distribution of $\nux$ is solely determined by the temperature, not the neutrino chemical potential. On the other hand, the contribution of the matter velocity is common between them. This difference explains the distribution of the $r\theta$ component of the Eddington tensors of $\nue$, $\nueb$, and $\nux$.
\end{enumerate}

We have conducted thorough analyses of the simulations with a nonrotating $11.2\,M_\odot$ progenitor with the LS and FS EOSs in this paper, but systematic studies about the input physics would be also required. We are further running the Boltzmann-radiation-hydrodynamics code with a variety of input physics: different progenitors, different initial rotations, and different EOSs. Collecting deep analyses of these simulations, we can obtain valuable clues for understanding the explosion mechanism of CCSNe.

In addition, we are continuously developing the code. For 2D simulations, improvement in the feedback from neutrino--matter interaction and its influence on the PNS kick are discussed in \citet{2019ApJ...878..160N, 2019ApJ...880L..28N}. These simulations employ one of the currently most realistic EOS models \citep{2017NuPhA.961...78T}. We have developed a 3D version of the Boltzmann-radiation-hydrodynamics code, and it can follow the first few tens of milliseconds \citep{2020arXiv200402091I}. A general relativistic version of the code is also under development. The spherical polar coordinate employed in our code has some difficulty at the center and along the pole when running the numerical relativity, but \citet{2013PhRvD..87d4026B} has suggested a method to avoid it. We are developing the general relativistic version of our code using their technique. With the coming next-generation supercomputers, we hope to run the general relativistic Boltzmann-radiation-hydrodynamics code without any spatial symmetry. The simulations with the code would provide deep understanding of the effect of each physical process on the dynamics of the SN explosions.

\acknowledgments
We thank K. Nakazato, Y. Suwa, and T. Takiwaki for fruitful discussions. This research used high-performance computing resources of the K-computer and the FX10 of the HPCI system provided by the R-CCS and the University of Tokyo through the HPCI System Research Project (Project ID: hp 140211, 150225, 160071, 160211, 170230, 170031, 170304, 180111, 180239), SR16000 and XC40 at YITP of Kyoto University, SR16000 and Blue Gene/Q at KEK under the support of its Large Scale Simulation Program (14/15-17, 15/16-08, 16/17-11), Research Center for Nuclear Physics (RCNP) at Osaka University, and the XC30 and the general common-use computer system at the Center for Computational Astrophysics, CfCA, the National Astronomical Observatory of Japan. The large-scale storage of numerical data is supported by JLDG constructed over SINET of NII. This work was supported in part by a Grant-in-Aid for Scientific Research on Innovative areas “Gravitational wave physics and astronomy: Genesis” (17H06357, 17H06365) from the Ministry of Education, Culture, Sports, Science and Technology (MEXT), Japan, and in part a Grant-in-Aid for Scientific Research (C; 15K05093, 19K03837, B; 20H01905) and Young Scientists (Start-up, JP19K23435) from the Japan Society for the Promotion of Science (JSPS). This work was also partly supported by research programs at K-computer of the RIKEN R-CCS, HPCI Strategic Program of Japanese MEXT, Priority Issue on Post-K-computer (Elucidation of the Fundamental Laws and Evolution of the Universe), Joint Institute for Computational Fundamental Sciences (JICFus), and by MEXT as “Program for Promoting Researches on the Supercomputer Fugaku” (Toward a unified view of the universe: from large scale structures to planets). A.H. was supported in part by the Advanced Leading Graduate Course for Photon Science (ALPS) in the University of Tokyo and a Grant-in-Aid for JSPS Research Fellows (JP17J04422). K.S. acknowledges the Particle, Nuclear and Astro Physics Simulation Program (2019-002, 2020-004) at KEK.

\software{matplotlib \citep{Hunter:2007}}



\bibliographystyle{aasjournal}
\bibliography{ref}

\end{document}